\documentclass[]{pasj01} 
\Received{$\langle$27--September--2021$\rangle$}
\Accepted{$\langle$10--April--2022$\rangle$}
\Published{$\langle$publication date$\rangle$}

\usepackage{longtable}
\usepackage{graphicx}
\usepackage{txfonts}
\usepackage{natbib}
\usepackage{subfloat}
\usepackage{lineno}
\def\la{\ifmmode{\lesssim}\else$\lesssim$\fi}
\def\ga{\ifmmode{\gtrsim}\else$\gtrsim$\fi}

\begin{document}

\title{W49N MCN-a: a disk-accreting massive protostar embedded in an early-phase hot molecular core}

\author{Ryosuke Miyawaki$^1$, Masahiko Hayashi$^{2,3}$, and Tetsuo Hasegawa$^2$}%

%
\altaffiltext{1}{College of Arts and Sciences, J.F. Oberlin University, Machida, Tokyo 194-0294, Japan}
\email{miyawaki@obirin.ac.jp}
\altaffiltext{2}{National Astronomical Observatory of Japan, 
2-21-1 Osawa, Mitaka, Tokyo 181-8588, Japan}
\altaffiltext{3}{JSPS Bonn Office, Ahrstr. 58, 53175 Bonn, Germany}


\maketitle

\begin{abstract}\label{ABSTRACT}

We present ALMA archival data for 219--235~GHz continuum and line observations toward the hot molecular core (HMC) W49N MCN-a (UCHII region J$_{1}$) at a resolution of $\sim$0\farcs3.
The dust continuum emission, showing an elongated structure of 1\farcs40$\,\times\,$0\farcs95 (PA=43.5\degree) perpendicular to the outflow seen in SiO and SO, represents a rotating flattened envelope, or torus, with a radius of 7,800~au inclined at 47.5\degree\ or larger.
The emissions from CH$_3$CN, $^{13}$CS, HNCO, HC$_3$N, SO$_2$, DCN, H$_2$CO, OCS,  CH$_3$OH, and C$^{18}$O exhibit a consistent velocity gradient as a result of rotation.
The magnitude of each velocity gradient is different, reflecting that each line samples a specific radial region.
This allows us to derive a rotation curve as $V_{\rm rot}\propto R^{0.44\pm0.11}$ for 2,400~au $\la\,R\,\la$ 14,000~au, giving the dynamical mass as 
$M_{\rm dyn} = 57.0^{+24.5}_{-17.1}\,(R\,[{\rm au}]/3,000)^{1.88}$\,M$_\odot$.
The envelope mass independently estimated from the dust emission is 910\,M$_\odot$ (for $T_{\rm dust}=180$~K)  for $R\le\,$7,800~au and 32\,M$_\odot$ (for $T_{\rm dust}=300$~K) for $R\le\,$1,700~au.
The dynamical mass formula agrees well with these mass estimates within an uncertainty of a factor of three in the latter.
The envelope is self-gravitating and is unstable to form spiral arms and fragments, allowing rapid accretion to the inner radii with a rate of order 10$^{-2}$\,M$_\odot$\,yr$^{-1}$, although inward motion was not  detected.
The envelope may become a non self-gravitating Keplerian disk at $R\,\la\,$~(300--1,000)~au.
The formula is also consistent with the total mass $\sim$10$^4$\,M$_\odot$ of the entire HMC 0.15~pc (31,000~au) in radius.
Multiple transitions of CH$_3$CN, HNCO and CH$_3$OH provide the rotation temperatures of 278\,$^{+39}_{-30}$~K, 297\,$^{+52}_{-39}$~K, 154\,$^{+73}_{-37}$~K, respectively, for $R\,\la\,$1,700~au, suggesting that the central source of MCN-a has an intrinsic bolometric luminosity of $\sim10^6$~L$_\odot$.
These results have revealed the structure and kinematics of MCN-a at its intermediate radii.
With no broad-line H30$\alpha$ emission detected, MCN-a may be in the earliest phase of massive star formation.

\end{abstract}

\section{Introduction}\label{INTRODUCTION}
Massive stars play key roles in the evolution of the Universe. 
Large dust extinction, in addition to the scarcity and short time scale of evolution, makes it difficult for us to observe the early phases of their formation.
Theoretical understanding of massive star formation is also arduous because of the complex physics involved.
The low number statistics of young or forming massive stars is only partially offset by their large luminosities, which allow us to study them at greater distances than their low-mass counterparts at the cost of spatial resolution.

Observational evidence has been accumulated that massive stars undergo the phases of hot molecular cores (HMCs) \citep{Kurtz2000, Cesaroni2005} and ultracompact HII (UCHII) regions.
HMCs are characterized by their compact sizes (\la\,0.1\,pc), large masses of warm ($T$\,\ga\,100\,K) and dense ($n({\rm H}_2)\,\ga\,10^7\,$ cm$^{-3}$) gas, and large abundances of complex organic molecules evaporated off dust grains. 
The HMC phase may last for 10$^4$--10$^5$~yr \citep{Herbst2009, Battersby2017}.

At the center of an HMC, an embedded massive star, or a group of stars, grows rapidly (\la10$^5$\,yr) by accretion, possibly through an equatorial disk and an associated outflow \citep{Tanaka2016}, and eventually ionizes the surrounding gas to form a UCHII region.
This stage corresponds to the hollow HMC \citep{Stephan2018},  the last stage of the HMC phase \citep[e.g.][]{Furuya2011, Rolffs2011, Serra2012, Fuente2018}.
The hollow HMC has the same density structure as the HMC, but it contains an ionized cavity at the center.  
The UCHII region expands, but stays confined to the stellar vicinity inside its surrounding massive envelope.
It takes $\sim$10$^5$~yr for a UCHII region to reach a radius of order a parsec and destroy the molecular core in which it was embedded \citep{Akeson1996, Churchwell2002, MacLow2007}.
In the later phases, the gas surrounding the massive stars is globally ionized, often by several ionizing sources, to form compact and then classical HII regions, disrupting the parent molecular cloud \citep{Yorke1986}. 

Hypercompact HII (HCHII) regions, much smaller in size than UCHII regions, may be considered to be in a transitional phase from an HMC to a UCHII region.
They have little or undetectable centimeter continuum emission and show rising flux densities toward millimeter wavelengths \citep{Kurtz 2005, Hoare2007}.
Hydrogen radio recombination lines detected from several HCHII regions exhibit very broad line widths (FWHM $\sim\,$50 to 180\,km\,s$^{-1}$) \citep{Kurtz 2005}.
A nearly edge-on dust disk is detected surrounding the HCHII region M17-UC1 \citep{Nielbock2007}.
These characteristics can be explained by models that HCHII regions are photoevaporating disks embedded in massive infalling envelopes \citep{Hollenbach1994, Lizano1996, Keto2007}.

Rotating structures around massive protostellar candidates are reported toward, e.g., IRAS18089-1732,  G24.78+0.08, G28.20-0.05, G31.41+0.31, and G10.6-0.4 \citep{Beltran2004, Beltran2005, Beuther2004, Beuther2005, Sollins2005a, Keto1987, Sollins2005b, Sollins2005c}.
These objects already have UCHII regions.
The rotating molecular clumps around them are large ($\sim$8,000--16,000~au in diameter) and massive ($\sim$80--500\,M$_\odot$).
The mass is not dominated by the central star, but by the surrounding gas, implying that the rotation is not Keplerian-like \citep{Beuther2004, Beuther2005}.
They may be self-gravitating envelopes or tori \citep{Beltran2005} and are distinguished from Keplerian-rotating disks \citep{Beltran2016}.
Their morphology and kinematics are not as well known as the latter.

Because of the high angular resolution available with ALMA, central regions of HMCs are partially resolved by recent observations.
The presence of a rotating disk around Orion Source I has long been known \citep[e.g.,][]{Kim2008}, but its detailed velocity structure has been revealed only recently with ALMA \citep{Hirota2017, Ginsburg2018}.
Presence of rotating disks has since been reported around other massive (proto-)stars, e.g., toward AFGL 4176 mm1, G29.96$-$0.02 HMC, G345.50+0.35~M and S, and G17.64+0.16 \citep{Johnston2015, Cesaroni2017, Maud2019, Zhang2019, Tanaka2020, Williams2022}.
Their velocity fields suggest possible Keplerian-like rotation at 100--1,000~au in radius.
Such a disk may be the very central part of an HMC.
It is not known how the gas in this inner, possibly Keplerian rotating disk is supplied from its surrounding larger structure in order for the central protostar to keep gaining mass to grow.

In this paper, we present observations of MCN-a, an HMC with a mass of 10$^4$\,M$_\odot$ in W49 North (hereafter called W49N) located at a distance of 11.11$^{+0.79}_{-0.69}$~kpc \citep{Zhang2013}.
MCN-a was first identified by \citet{Wilner2001} as a compact source of methylcyanide (CH$_3$CN) emission coincident with a dust continuum source that they named K2.   
While other CH$_3$CN HMCs in W49N are located toward or in the vicinity of the prominent UCHII regions, MCN-a is rather isolated.
It coincides with the inconspicuous UCHII region J$_{1}$ \citep{DePree1997, Miyawaki2022}, whose Lyman continuum luminosity corresponds to a B0 star.
Far-infrared observations at 20~$\mu$m and 37~$\mu$m with {\it SOFIA} barely separated the two UCHII regions J$_{1}$ and J$_{2}$ \citep{DeBuizer2021}. 
Although J$_{2}$ is brighter at 20~$\mu$m, J$_{1}$ becomes brighter at 37~$\mu$m, suggesting that J$_{1}$ is the more embedded of the two sources. The observed bolometric luminosity of J$_{1}$+J$_{2}$ is 2.93$\times 10^{4}$~L$_\odot$, but the extinction-corrected bolometric luminosity could be as large as 2$\times 10^6$~ L$_\odot$ as a result of the large visual extinction of $A_{\rm v}$=233~mag toward J$_{1}$+J$_{2}$ \citep{DeBuizer2021}.
No masers (CH$_3$OH class I and II, H$_2$O, SiO, and OH) are directly associated with MCN-a \citep{Hu2016, Beuther2019, Ladeyschikov2019, Phetra2021}, implying that it is in a very early stage of massive star formation  \citep[e.g.,][]{Miyawaki2021}. 
We present high resolution images of MCN-a in the 226\,GHz continuum and various ``hot core'' molecular lines and discuss its structure and velocity field.
We assume that the source names MCN-a (CH$_{3}$CN hot core), K2 (millimeter-wave dust continuum source), and J$_{1}$ (centimeter-wave thermal free-free continuum source) represent the same object in this paper and use the name MCN-a for representing all these emission sources.

\section{ALMA archival data}\label{OBSERVATIONS}

We used Atacama Large Millimeter/submillimeter Array (ALMA) archival data (\#2016.1.00620.S: PI A. Ginsburg) for the study of HMCs in W49N. 
The observations were performed from July 2017 to September 2018 using 43--45 12-m antennas and from April to July 2017 using 9--12 7-m antennas of Morita array. 
The shortest and longest baselines for the 12-m antennas were 15.0\,m and 3,696.9\,m, respectively.
Those for the 7-m antennas were 8.9\,m and 48.9\,m, respectively.
The two datasets obtained with 12-m antennas and 7-m antennas were concatenated, while the total power mode data was not used.
Flux, bandpass, pointing, and phase calibrations were carried out with J1905+0952, J1924-2914, and J1922+1530, J1830+0619. 

Four spectral windows (216.90--218.90 GHz, 218.85--220.85~GHz, 230.86--232.86~GHz, and 232.73--234.73~GHz), each covering a $~2$\,GHz bandwidth, were set up to observe the target source W49N.
Image analysis was done using the CASA software \citep{McMullin2007}.
We separated the continuum and line emissions by fitting a linear baseline to line free channels of each spectral window using `uvcontsub' task of CASA.
The continuum data  at the effective frequency of 226\,GHz was then made by averaging the line free intensities of all the four windows.

For each spectral line, we set up a data cube with a spectral resolution of 2~kms$^{-1}$.
Although the frequency resolution (channel width) was 976.56\,kHz ($\sim$1.4 km\,s$^{-1}$), there was variation in frequency to channel relation between datasets obtained on different days.
We thus rounded the flux received in an original spectral channel into the nearest 2~kms$^{-1}$ bin in order to compensate the variation.

The phase center was $\alpha$(ICRS) = 19$^{\rm h}$10${\rm ^m}$13\,\fs\,5, $\delta$(ICRS) =09\degree06$'$12$\,\farcs$00. 
We set the parameters of `tclean' task as weighting=`briggs' and robust=`0.5'.
This setting creates a PSF that smoothly varies between natural and uniform weighting based on the signal-to-noise ratio of the measurements and a tunable parameter that defines the noise threshold.
The synthesized beam size of the continuum image was 0\farcs31 $\times$ 0\farcs26 (PA=$-67.3\degree$).
The synthesized beam sizes of the line images were 0\farcs35$\times$ 0\farcs34 (PA=$-87\degree$) and 0\farcs35 $\times$ 0\farcs29 (PA=$-68\degree$) at 219\,GHz and 233\,GHz, respectively.
The resultant noise levels were 1.4\,mJy\,beam$^{-1}$ and 4.3\,mJy\,beam$^{-1}$ for the continuum and line maps, respectively.

\section{Results}\label{RESULTS}

\subsection{Continuum emission}\label{Results-Cont}

\begin{figure}[htbp]
\includegraphics[bb= 100 200 450 680, scale=0.55]{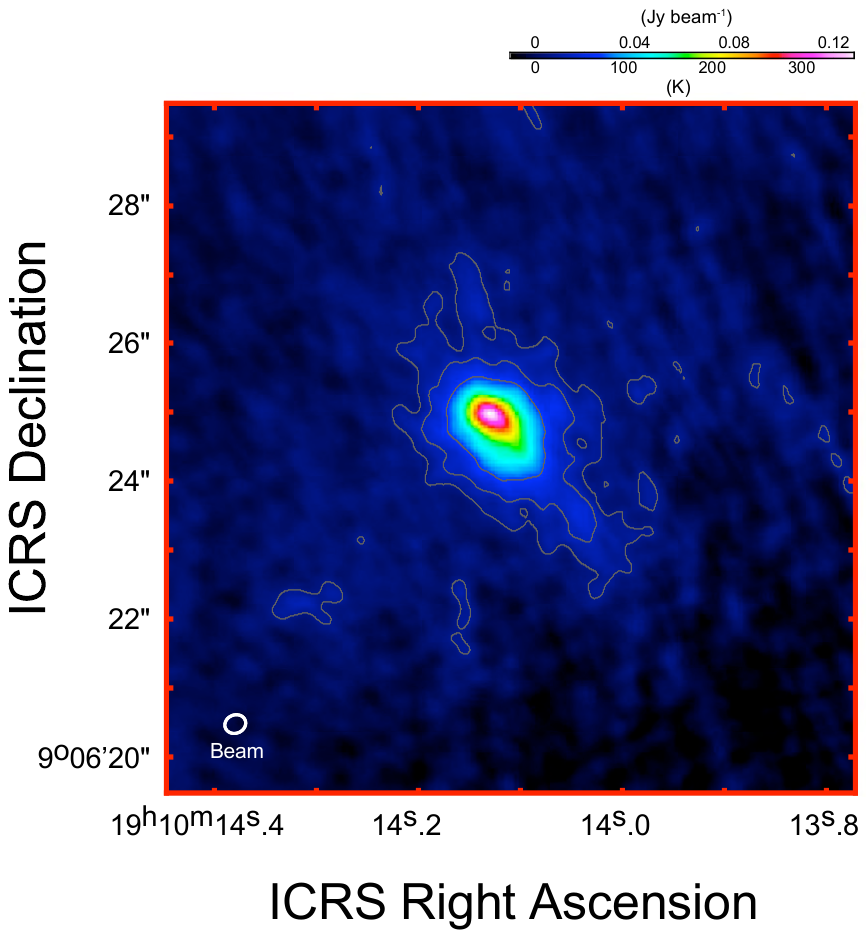}
\caption{226\,GHz continuum map. Contours are drawn at 5, 10, and 20\% levels of the peak brightness.
}
\label{Fig01}
\end{figure}

Figure~\ref{Fig01} shows the 226\,GHz continuum image toward W49N MCN-a.
It is elongated in the NE to SW direction (PA=43.5$^\circ$).
Applying the CASA `Two-Dimensional Fitting' tool to the area above the 5\% level of the peak, we measured the peak brightness, total flux density, and deconvolved size to be 128.5\,mJy\,beam$^{-1}$, 1,462$\,\pm\,$50\,mJy, and 1\farcs40$\,\times\,$0\farcs95 (FWHM), respectively.

\begin{figure}[htbp]
\includegraphics[bb= 30 40 500 520, scale=0.5]{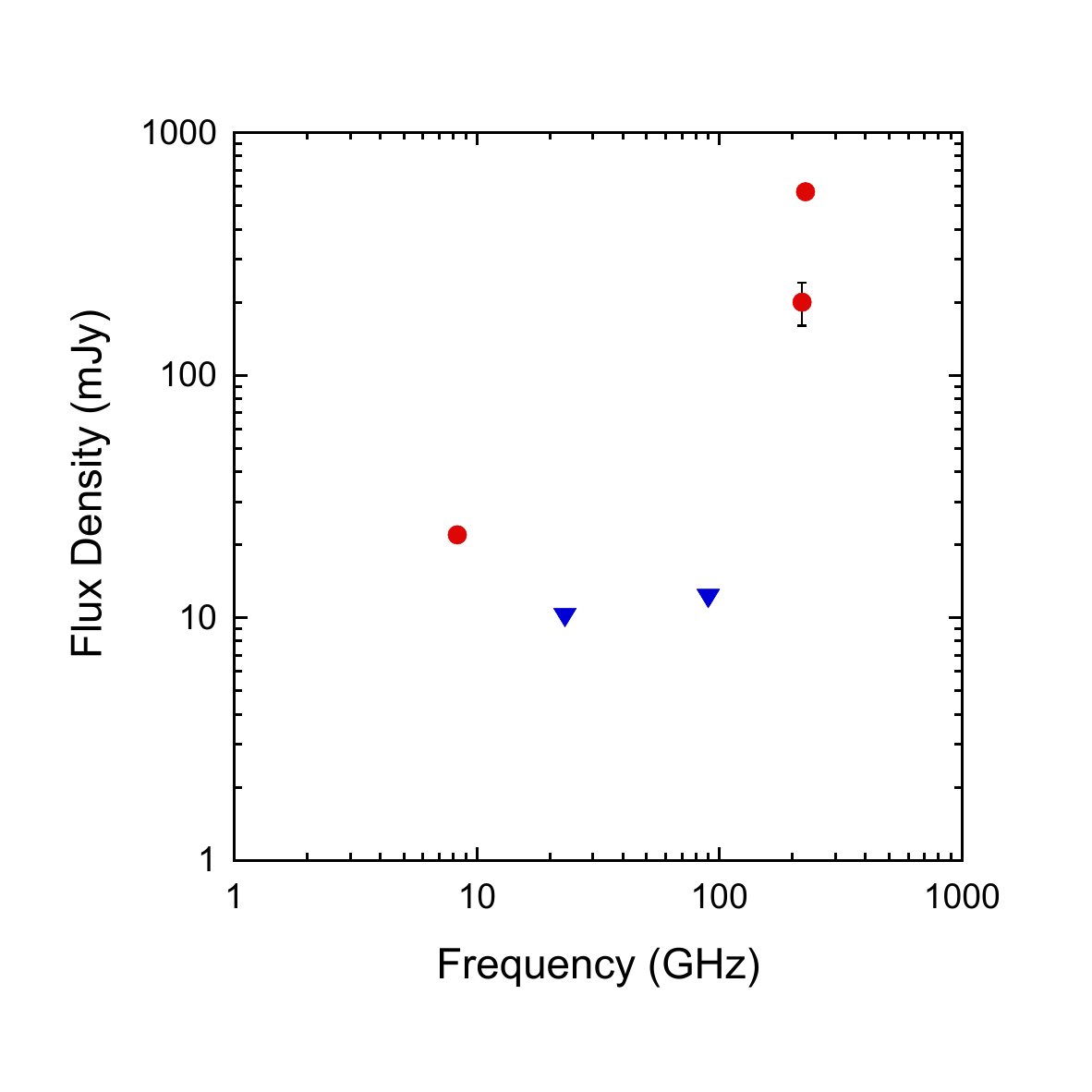}
\caption{Flux density distribution of MCN-a. 
Red circles show the flux densities detected, while blue triangles are upper limits.
Note that the flux densities at 219\,GHz and 226\,GHz are measured in a 1$''$ box and a 1$''$ diameter beam, respectively, toward the peak.
Error bars are shown when larger than the plot circles.
}
\label{Fig02}
\end{figure}

The millimeter-wave continuum source was previously detected at 219\,GHz with the Berkeley-Illinois-Maryland Association millimeter-wave interferometer (BIMA) \citep{Wilner2001}, also showing an elongated structure from NE to SW with a 0\farcs18 restored beam.
The 219\,GHz flux density was 200$\,\pm\,$40\,mJy in a 1$''$ box toward the peak.
The source was not detected at 90\,GHz with an upper limit of $<$12\,mJy (3$\sigma$) \citep{DePree2000}.
The 219\,GHz flux density and the 90~GHz upper limit give a spectral index $\alpha$ ($F_\nu\propto\nu^\alpha$) larger than 3.0, indicating that the 219\,GHz emission originates from dust particles \citep{Wilner2001}.
The current data at 226\,GHz has the flux density of 570\,mJy within a 1$''$ diameter beam, which confirms that the emission at this band originates from dust particles.
For reference, the thermal free-free emission from MCN-a has a flux density of 22\,mJy at 8.3\,GHz \citep{DePree1997} and was not detected at 23\,GHz with an upper limit of $\la$10\,mJy \citep{DePree1997}.
A flux density distribution plot in the radio and millimeter wavelengths is shown in Figure~\ref{Fig02}.

The elongated dust emission of MCN-a provides strong support for a tilted disk-like structure, at the center of which is located a (proto)star emitting Lyman continuum photons equivalent to a B0 star \citep{DePree1997}.
Although we cannot exclude the possibility that the elongation is caused by other dust emission sources aligned on the major axis, systematic velocity gradients along the axis traced by various molecular lines and a bipolar outflow perpendicular to it, to be presented in the next subsection, are clear manifestation of the disk-like structure.
Its radius, i.e., the length of the semi-major axis, is 7,800\,au with the assumed distance of 11.11~kpc to W49N \citep{Zhang2013}.
Assuming a geometrically thin disk, we obtain the inclination angle of 47.5\degree\,$\pm$\,2.4\degree\ (0\degree\ for face-on) from the minor to major axis ratio.
This should be a lower limit because the structure may be geometrically thick, as will be discussed in the following sections.
The structure is also massive and self-gravitating as we see in \S\ref{DISCUSSION}.
We will thus call it as a flattened and/or rotating envelope hereafter because it is much different from the Keplerian rotating disks around low mass young stars.
Table~\ref{Table01} summarizes the results of continuum emission.

\begin{table}[ht]
\caption{Summary of the 226\,GHz Continuum Observations}
\label{Table01}
\begin{center}
\scalebox{0.7}[0.7]

\begin{tabular}{llll}

\hline\hline
\multicolumn{3}{l}{Peak Position}\\
\ \ $\alpha$(ICRS)& 19$^{\rm h}$10${\rm ^m}$14\fs123 & $\pm$0\fs001 \\
\ \ $\delta$(ICRS)& 09\degree06$'$24$\farcs$84 & $\pm$0$\farcs$02 \\
\ \ Brightness & 128.5 mJy\,beam$^{-1}$ & $\pm$1.4 mJy\,beam$^{-1}$ \\
\multicolumn{3}{l}{Flux Density}\\
\ \ Total Integrated & 1,462 mJy & $\pm$50 mJy \\
\ \ In $\Omega_{\rm beam}$\footnotemark[*] & 84.3 mJy & $\pm$2.7 mJy \\
\multicolumn{3}{l}{Emission Size}\\
\ \ Major Axis\footnotemark[$\dagger$]& 1$\farcs$40 & $\pm$0$\farcs$05 \\
\ \ Minor Axis\footnotemark[$\dagger$] & 0$\farcs$95 & $\pm$0$\farcs$03 \\
\ \ Position Angle& 43.5\degree & $\pm$3.5\degree \\
\multicolumn{3}{l}{Flattened Envelope Model}\\
\ \ Radius& 7,800 au & $\pm$270 au \\
\ \ Inclination\footnotemark[$\ddagger$] &  47.5\degree & $\pm$2.4\degree \\
\hline
\end{tabular}
\end{center}
\footnotetext{1}{$^*$Integrated over the beam solid angle toward the peak position}\\
\footnotetext{1}{$^\dagger$Beam deconvolved FWHM}\\
\footnotetext{2}{$^\ddagger$Assuming a geometrically thin disk}
\end{table}


\subsection{Line emission}\label{Results-Line}

\begin{figure*}[htbp]
\includegraphics[bb= 110 330 700 760, scale=1.0]{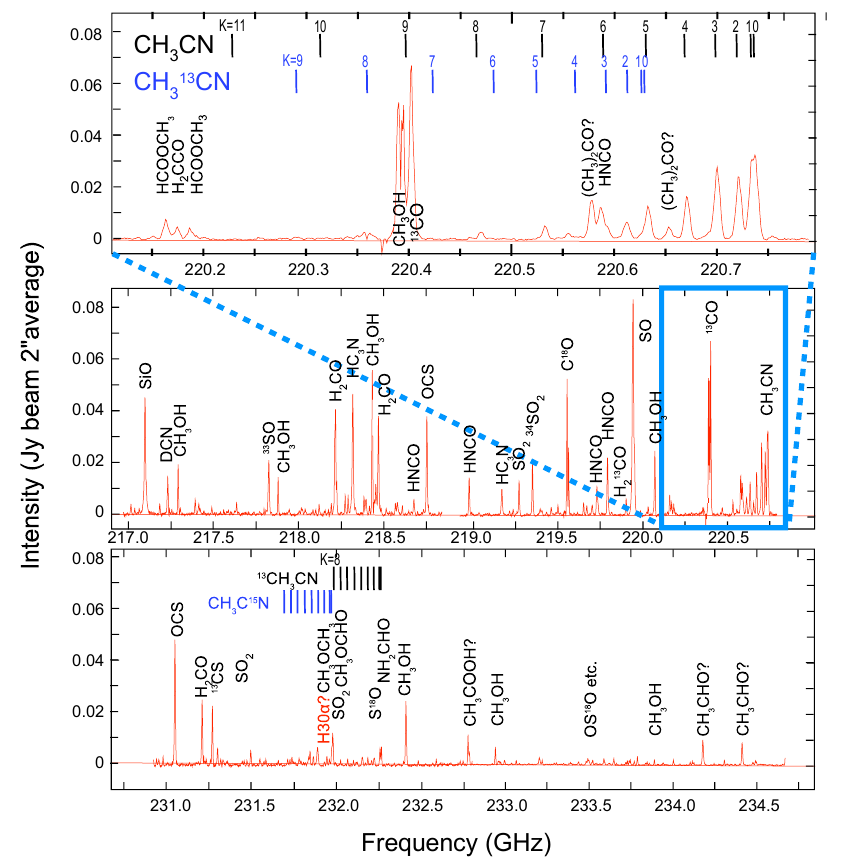}
 \caption{Emission lines in the four spectral windows averaged over a 2$''$ diameter beam after the continuum emission is subtracted.
Upper panel: Enlargement of the spectrum in the blue rectangle of the mid panel.
Rotational lines of CH$_3$CN ($J_K=12_K-11_K$) and CH$_3$$^{13}$CN ($J_K=12_K-11_K$) for various $K$-states are identified.
Mid panel: Spectral windows of 216.90--218.90 GHz and 218.85--220.85~GHz.
Lower panel: Spectral windows of 230.86--232.86~GHz and 232.73--234.73~GHz.
Frequencies for the rotational lines of $^{13}$CH$_3$CN ($J_K =13_K-12_K$) and CH$_3$C$^{15}$N ($J_K =12_K-11_K$) for various $K$-states are indicated.
\label{Fig03}}
\end{figure*}

Figure~\ref{Fig03} shows the emission lines in the entire range of the four spectral windows after the continuum emission is subtracted.
The power spectrum is averaged over a beam of 2$''$ in diameter.
Various transitions of molecular lines were detected and are identified in the figure.
The hydrogen recombination line H30$\alpha$ was not detected (see Appendix~\ref{U-line}).
An enlargement of the higher frequency part of the mid panel is shown in the upper panel, where $J_K=12_K-11_K$ transitions with different $K$ values of CH$_3$CN and CH{$_3$}$^{13}$CN are identified.

\begin{figure*}[htbp]
\includegraphics[bb= 0 210 700 610, scale=0.75]{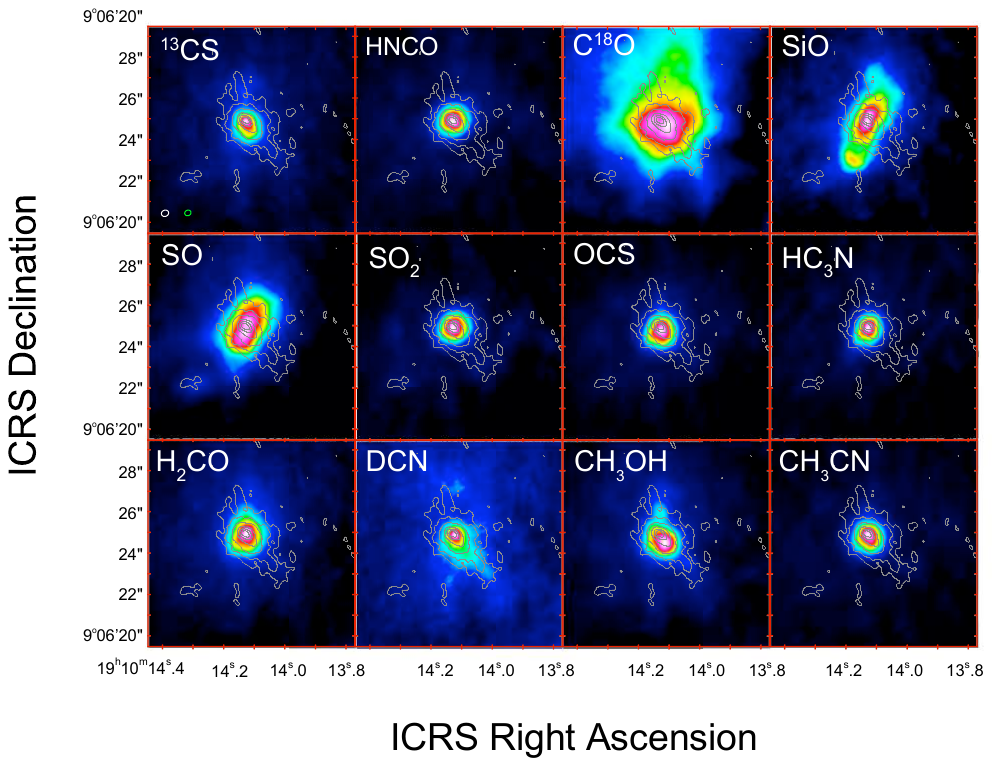}
\caption{Integrated intensity (moment 0) maps superimposed on the contours of the continuum emission at 5, 10, 20, 40, 60, and 80\% levels of the peak brightness.
Color scales are shown above the velocity channel maps that follow.
\label{Fig04}}
\end{figure*}

Figure~\ref{Fig04} shows the integrated intensity maps of $^{13}$CS ($J=5-4$), HNCO ($J_{K_a,\,K_c}=10_{\,0,\,10}-9_{\,0,\,9}$), C$^{18}$O ($J=2-1$), SiO ($J=5-4$), SO ($N_J=6_5-5_4$), SO$_{2}$ ($J_{K_a, K_c}=28_{3,25}-28_{2,26}$), OCS ($J=19-18$), HC$_{3}$N ($J=24-23$), H$_{2}$CO ($J_K=3_{22}-2_{21}$), DCN ($J=5-4$), CH$_{3}$OH ($J_K=4_{3}-3_{2}$) and CH$_{3}$CN ($J_3=12_3-11_3$).
We notice that the maps of $^{13}$CS, HNCO, SO$_{2}$, OCS, HC$_{3}$N, H$_{2}$CO, DCN, CH$_{3}$OH, and CH$_{3}$CN are more or less circular and less extended than the dust emission.
We will call them as the compact emission molecules.
The C$^{18}$O map is, on the other hand, much more extended.
The SiO and SO maps are elongated nearly perpendicular to the dust emission.
The peak brightness, peak integrated intensity (see \S\ref{Trot}), and minor to major axis ratio for each of these maps are listed in Table~\ref{Table02}.

\begin{figure*}[htbp]
\includegraphics[bb= 0 210 700 610, scale=0.75]{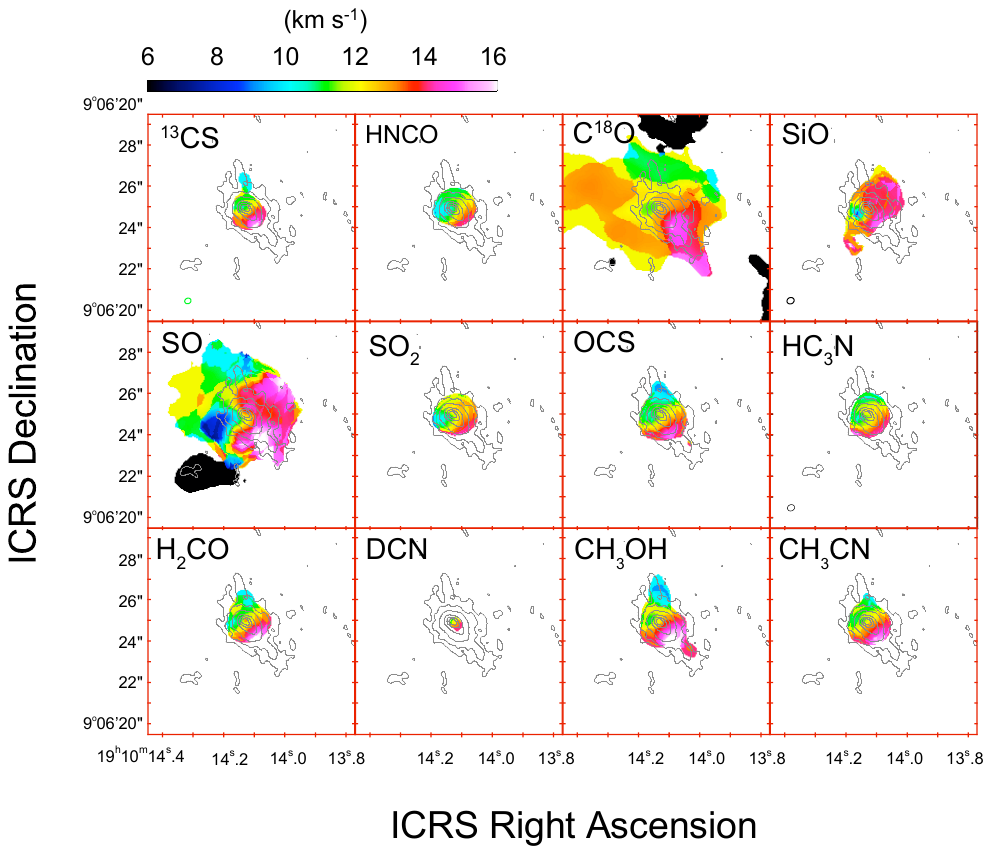}
\caption{Mean velocity (moment 1) maps superimposed on the contours of the continuum emission at 5, 10, 20, 40, 60, and 80\% levels of the peak brightness.
The color-coded mean velocity is shown for the area above the 80\,mJy\,beam$^{-1}$ level of the peak brightness of each line emission.
The intensity weighted velocity is integrated from $V_{\rm LSR}=$ 0 to 25 \,km\,s$^{-1}$ except for C$^{18}$O, for which the integration range was from 5 to 25\,km\,s$^{-1}$ to exclude the background emission around $V_{\rm LSR}=$ 0\,km\,s$^{-1}$.
\label{Fig05}}
\end{figure*}

Figure~\ref{Fig05} shows the mean velocity (moment~1) maps for the lines shown in Figure~\ref{Fig04}.
The compact emission molecules exhibit a clear velocity gradient along the major axis of the flattened envelope: the lines of equal velocity are overall perpendicular to the elongation of the dust emission.
The position angle of the rotating axis is thus well defined regardless of the envelope geometry. 
The C$^{18}$O map appears to have a complicated velocity field, but show some hint of a velocity gradient from NE to SW, about which we will discuss later (\S\ref{C18O}).
The SiO and SO emissions have a major velocity gradient along their elongations, i.e., from SE to NW perpendicular to the flattened envelope.
This overall velocity gradient is caused by the outflow, as will be discussed in \S\ref{SiO} and \S\ref{SO}.
It suggests that the redshifted outflow is on the NW side and blueshifted outflow is on the SE side.
On the SE side of the elongated dust emission, the SiO and SO emissions appear to have a velocity gradient also from NE to SW.
This gradient may partly reflect the rotation of the flattened envelope for SO, but may not do so for SiO, as will be examined in detail in \S\ref{SiO} and \S\ref{SO}.

\subsubsection{CH$_3$CN emission}\label{CH3CN}

\begin{figure*}[htbp]
\includegraphics*[bb= 0 210 700 610, scale=0.75]{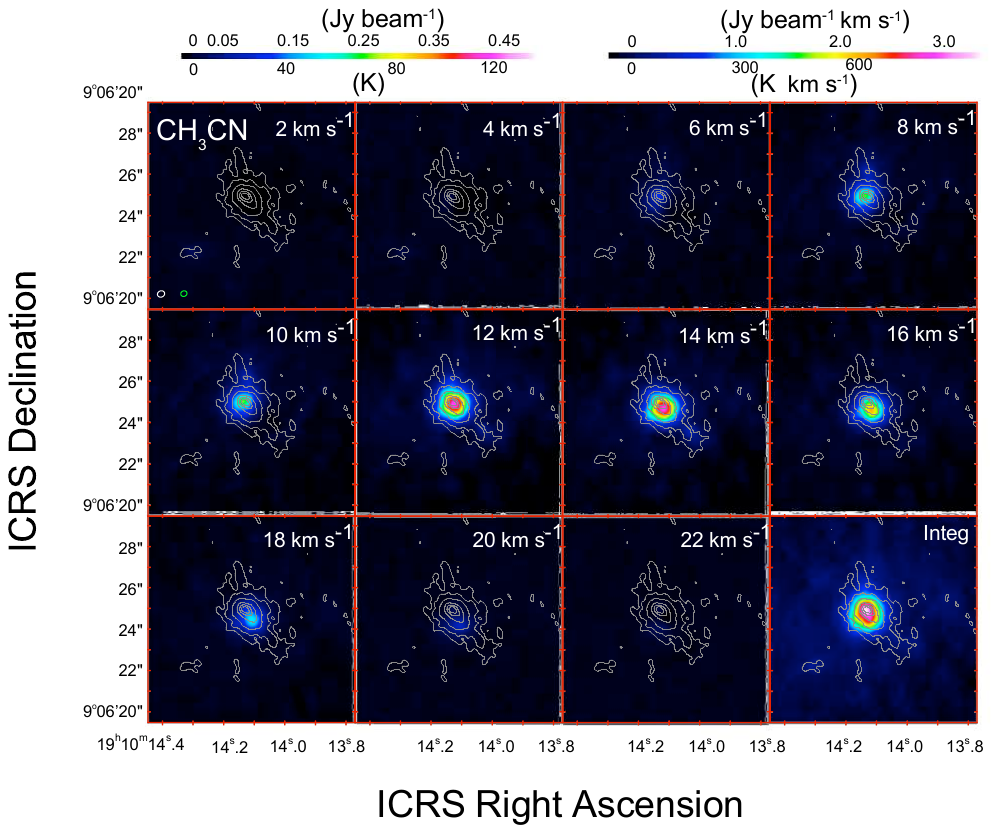}
\caption{Velocity channel maps of the CH$_3$CN ($J_K=12_3-11_3$) line superimposed on the contours of the continuum emission at 5, 10, 20, 40, 60, and 80\% levels of the peak brightness.
Bottom-right panel shows the total integrated intensity taken from Figure~\ref{Fig04}.
The color bars above the maps show the intensity scales for the channel maps (left) and for the integrated intensity map (right).
\label{Fig06}}
\end{figure*}

Figure~\ref{Fig06} shows the velocity channel maps of the CH$_3$CN ($J_K=12_3-11_3$) line, which is the strongest unblended emission line of CH$_3$CN (see the upper panel of Figure~\ref{Fig03}).
For comparison, contours of the continuum emission are drawn at the  5, 10, 20, 40, 60, and 80\% levels of the peak brightness.
The CH$_3$CN emission is significantly detected at velocities from $V_{\rm LSR}$ =\,6\, km\,s$^{-1}$ to 20\, km\,s$^{-1}$.
The dominant part of the emission at each velocity originates inside the 20\% level contour of the continuum emission representing the tilted flattened envelope.

As the radial velocity increases from $V_{\rm LSR}$ =\,8\, km\,s$^{-1}$ to 16\, km\,s$^{-1}$, the peak position moves from the NE to SW across the continuum peak, with the line and continuum peaks coinciding with each other at 12\, km\,s$^{-1}$.
The velocity gradient demonstrates the rotation of the envelope, where the NE side is approaching us and the SW side is receding from us.

The channel maps of $^{13}$CS (J=5-4), HNCO ($J_{K_a,\,K_c}=10_{\,0,\,10}-9_{\,0,\,9}$), HC$_3$N ($J=24-23$), SO$_2$ ($J_{K_a,K_c}=28_{\,3,25}-28_{\,2,26}$), DCN ($J=5-4$), H$_2$CO ($J_K=3_{22}-2_{21}$), OCS ($J=19-18$), and CH$_3$OH ($J_K=4_{3}-3_{2}$) lines basically show the same tendency as the CH$_3$CN line.
Description of these line maps is given in Appendix~\ref{Channel-maps}.

\subsubsection{SiO emission}\label{SiO}

\begin{figure*}[htbp]
\includegraphics*[bb= 0 100 700 840, scale=0.75]{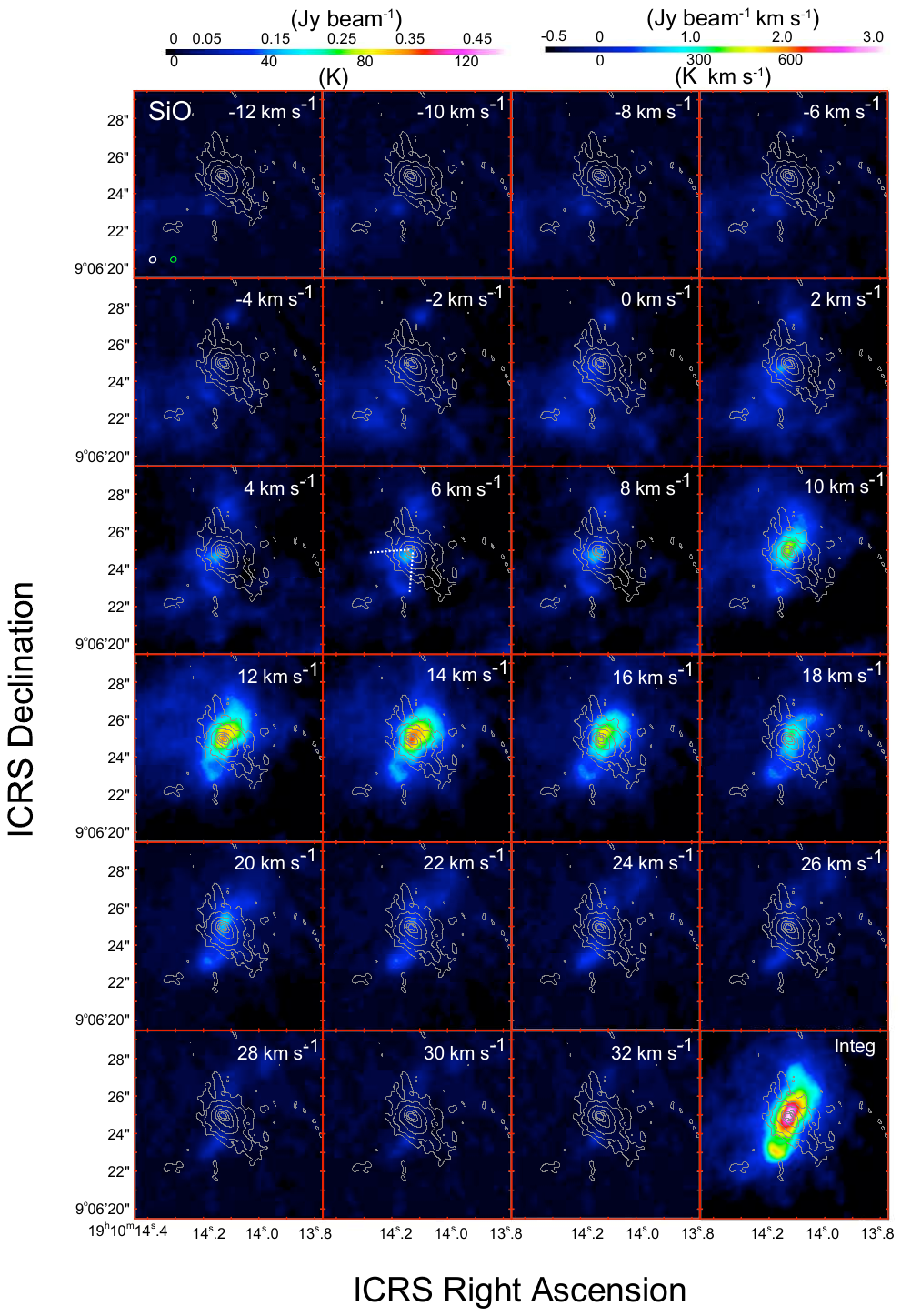}
\caption{Same as Fig.~\ref{Fig06}, but for the SiO ($J=5-4$) line.
Two white dotted lines on the 6\,km\,s$^{-1}$ panel indicate the possible opening angle of the blueshifted outflow.
\label{Fig07}}
\end{figure*}

Figure~\ref{Fig07} shows the velocity channel maps of the SiO ($J=5-4$) line.
Elongated features perpendicular to the continuum emission are seen at velocities from $V_{\rm LSR}=$\,10\,km\,s$^{-1}$ to 26\, km\,s$^{-1}$, with the emission ridges always passing through the continuum peak.
The geometrical relation suggests that the SiO emission represents an outflow emanating from the vicinity of the central star, as is consistent with the generally accepted idea that SiO emission originates in shocks caused by outflows \citep[e.g.,][]{Ziurys1989, Hirota2017}. 

Because the emission ridges always pass through the continuum peak, we do not find a clear, systematic shift in the ridge positions across the continuum peak around the systemic velocity of $V_{\rm LSR}=$10--18\, km\,s$^{-1}$, where contribution from the envelope emission could be expected.
This suggests that, even toward the center of the rotating envelope, the SiO emission originates mainly from the outflow, but not from the rotating structure.

On the NW side of the continuum peak, the SiO emission always extends to the NW at $V_{\rm LSR}=$10--18\, km\,s$^{-1}$.
On the SE side, however, the elongated SiO emission changes its direction from the SE to S as the velocity increases from 10 to 18\,km\,s$^{-1}$, resulting in the SiO ridge relatively straight at 10\, km\,s$^{-1}$, but becoming a little wiggled at 18\,km\,s$^{-1}$.
This gives rise to the apparent velocity gradient parallel to the elongated dust emission on its SE side seen in the mean velocity map of SiO (Figure~\ref{Fig05}).
The velocity gradient parallel to the dust emission, as a consequence, does not mean the rotation of the outflow about its axis either, as was observed toward Orion Source~I \citep{Hirota2017}.
We examined whether there is a systematic velocity gradient  perpendicular to the outflow axis at its various locations, but did not find a consistent trend suggestive of outflow rotation.

There is a weak but definite emission feature seen at $\sim$0\farcs5 SE at each of the blueshifted velocity channels of $V_{\rm LSR}=$2--8\, km\,s$^{-1}$.
It looks like having a fan shape with an apex at the continuum peak and an opening toward the SE.
It has an opening angle of $\sim$80\degree\ as is indicated by the two white dotted lines on the 6\,km\,s$^{-1}$ panel.
A similar fan-shaped outflow feature is also seen in SO (see the next subsection).
A possible counterpart feature on the redshifted side is seen on the 20\,km\,s$^{-1}$ panel, extending to the NNW from the continuum peak overlapped with faint emission elongated from the SSE to NNW.
It also has a fan shape with an opening angle of $\sim$40\degree.
At the more redshifted velocities $V_{\rm LSR}=$22--28\, km\,s$^{-1}$, we see faint, elongated emission passing through the continuum peak.
These characteristics of SiO emission suggest that the SE side of the outflow is mainly approaching us, while the NW side is receding from us near the outflow origin, implying that the NW side of the flattened envelope is the near side.

The presence of both blueshifted and redshifted emissions on both SE and NW sides of the origin means that the outflow may be ejected closer to the plane of the sky, namely the inclination may be larger than 47.5\degree.
If we take the roughly measured opening angles ($\sim$40\degree to 80\degree) of the fan-shaped blueshifted and redshifted features as that of the outflow, the outflow axis should be inclined by $\sim30\degree\pm10$\degree\ with respect to the plane of the sky in order for both blueshifted and redshifted emissions to be detected on both sides of its origin.
This means that the flattened envelope might have an inclination angle of 60\degree$\pm$10\degree\ with respect to the line of sight.
It also implies that the envelope is geometrically thick.
This inclination estimate is, however, very rough with a large uncertainty, so we will use 47.5\degree\ in this paper.
The inclination angle of 60\degree\ would reduce the dynamical mass that we will derive in \S\ref{Mass Distribution} by $\sim$30\%.

\subsubsection{SO emission}\label{SO}

\begin{figure*}[htbp]
\includegraphics*[bb= 0 100 700 840, scale=0.75]{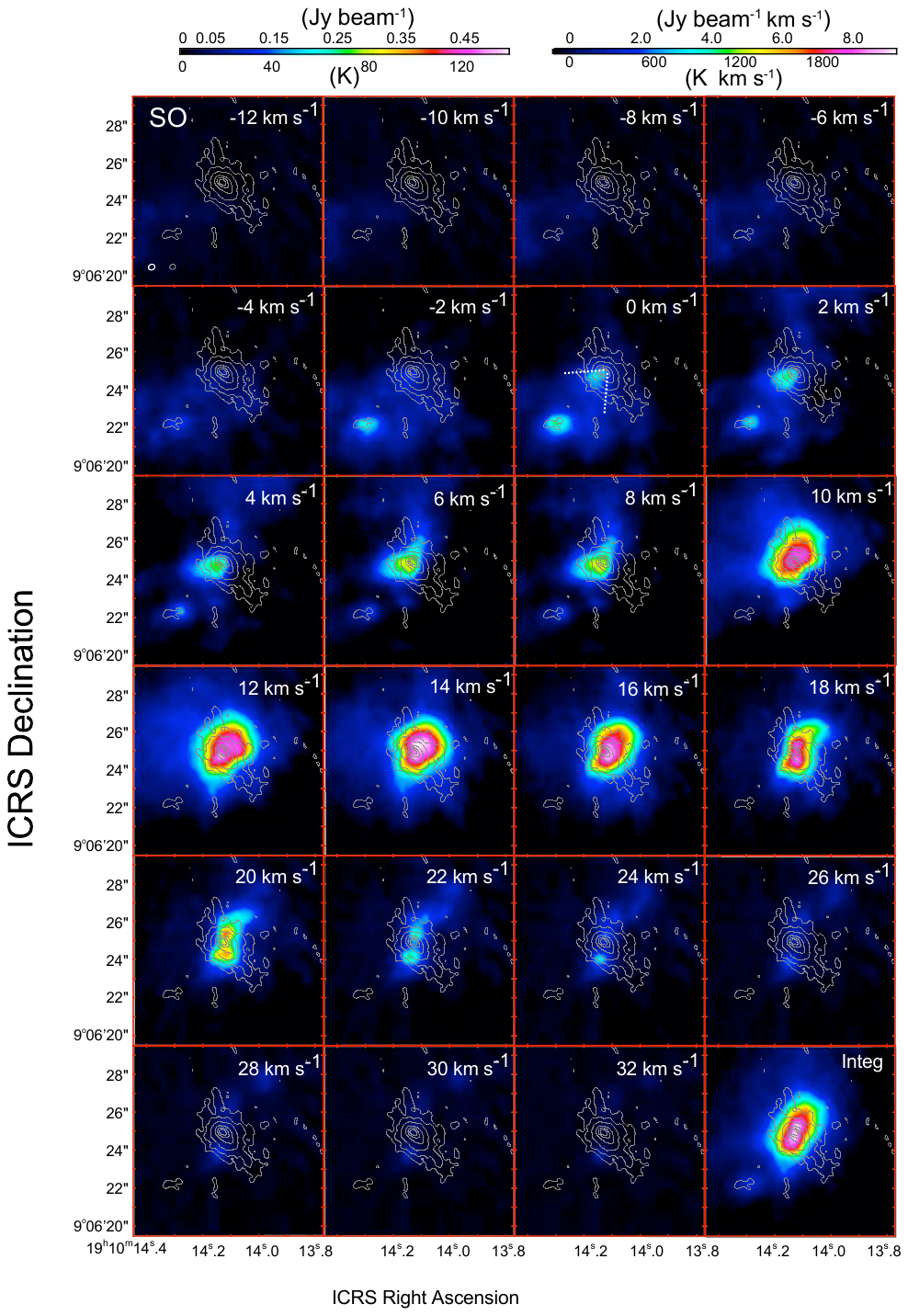}
\caption{Same as Fig.~\ref{Fig06}, but for the SO ($N_J=6_5-5_4$) line.
Two white dotted lines on the 0\,km\,s$^{-1}$ panel indicate the possible opening angle of the blueshifted outflow.
\label{Fig08}}
\end{figure*}

Figure~\ref{Fig08} shows the velocity channel maps of the SO ($N_J=6_5-5_4$) line.
The integrated intensity image on the bottom-right panel exhibits elongated emission nearly perpendicular to the flattened envelope.
This means that the emission arises from the outflow in large part. 
The emission ridges in the channel maps at higher  ($\ge$22\,km\,s$^{-1}$) and lower ($\le$4\,km\,s$^{-1}$) velocities pass through the continuum peak.
Similar to the SiO emission, a fan-shaped feature is seen at the blueshifted velocities of $V_{\rm LSR}=$0--8\, km\,s$^{-1}$ extending to the SE from its apex coincident with the continuum peak.
It also subtends an opening angle of $\sim$80\degree, similar to the corresponding SiO feature, as is indicated by the two white dotted lines on the 0\,km\,s$^{-1}$ panel.
The redshifted counterpart of this feature may be best seen at $V_{\rm LSR} =$22\, km\,s$^{-1}$ and is also visible at 20\,km\,s$^{-1}$ and 24\,km\,s$^{-1}$ at $\sim$0\farcs5 NNW of the continuum peak.
There is an additional redshifted feature at the south of the continuum peak at these velocities.
At $V_{\rm LSR}=$10--20\,km\,s$^{-1}$, the emission is elongated from the SE to NW with the strongest emission more or less shifted to the NW with respect to the continuum peak.
These variations of emission with velocity, namely the blueshifted outflow at the SE of the continuum peak and the redshifted outflow at its NW, resulted in the overall velocity gradient from SE to NW seen in Figure~\ref{Fig05}.

When we take a closer look at the channel maps around the systemic velocity ($V_{\rm LSR} \sim$ 12\,km\,s$^{-1}$), the emission ridge is slightly shifted to the NE of the continuum peak at 10\,km\,s$^{-1}$, while it moves to the SW at redshifted velocities of 14--16\,km\,s$^{-1}$.
The tendency is better seen in the SO emission arising from within the 40\% level contour of the continuum emission.
This shift of emission ridges with velocity across the continuum peak suggests that part of the SO emission toward the continuum peak originates in the rotating envelope.
Again, similar to the case of SiO, we do not think this shift arises from the rotation of the outflow because we did not find a consistent trend of velocity gradient perpendicular to the outflow axis at its various locations.

\subsubsection{C$^{18}$O emission}\label{C18O}
\begin{figure*}[htbp]
\includegraphics*[bb= 0 100 700 840, scale=0.75]{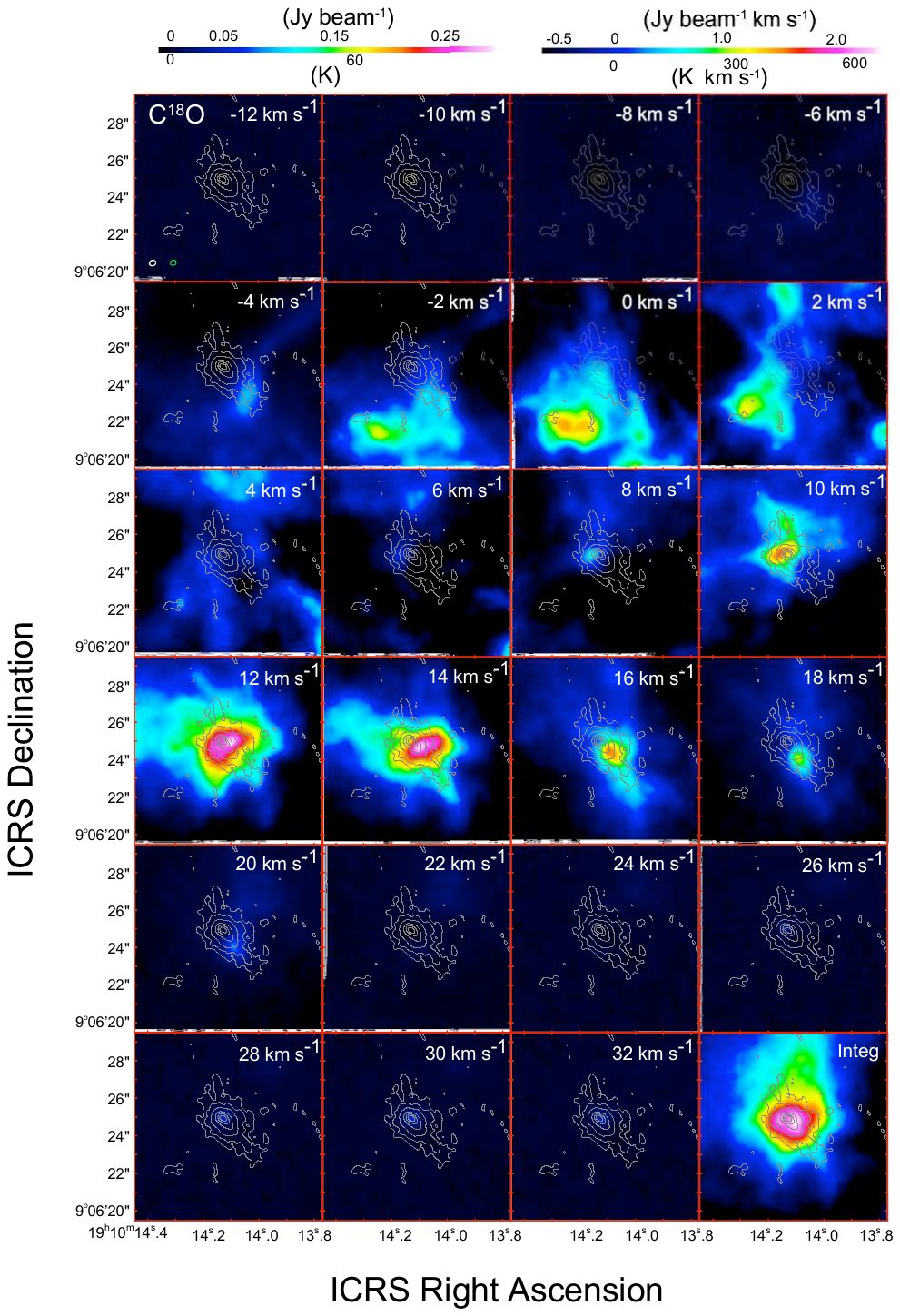}
\caption{Same as Fig.~\ref{Fig06}, but for the C$^{18}$O ($J=2-1$) line.
\label{Fig09}}
\end{figure*}

Figure~\ref{Fig09} shows the velocity channel maps of the C$^{18}$O ($J=2-1$) line.
Although the emission is contaminated by ambient gas especially at blueshifted velocities ($V_{\rm LSR}=-$4 to 4\,km\,s$^{-1}$), the gas associated with the dust emission is unambiguously identified at $V_{\rm LSR}\ge$ 8\,km\,s$^{-1}$.

We can see in the channel maps that the C$^{18}$O emission also exhibits rotation.
The emission appears at the E to NE of the continuum peak at $V_{\rm LSR}$=8\,km\,s$^{-1}$, becoming larger and elongated perpendicular to the flattened envelope at 10\,km\,s$^{-1}$.
The emission is located closest ($\sim$0\farcs5 NW) to the continuum peak at 12\,km\,s$^{-1}$, where the emitting area is larger than the entire dust continuum emission.
The emission has its peak shifting to the SW of the dust emission at 14\,km\,s$^{-1}$, again elongated perpendicular to the envelope.
At 16--18\,km\,s$^{-1}$, the emission becomes circular and smaller at the SW of the envelope.

The integrated intensity map of C$^{18}$O (bottom-right panel of Figure~\ref{Fig09}) has a FWHM size of 7\farcs00\,$\times$\,4\farcs44 (PA=173\degree), or 0.19~pc and 0.12~pc in semi-major and semi-minor axes, respectively.
The size is essentially the same as the isolated HMC SiO-NE \citep{Miyawaki2022}, which has the radius and mass of 0.15~pc and $(1-2)\times10^4$~M$_\odot$, respectively.
The mass derived from the current C$^{18}$O data ($(3.8-8.2)\times10^3$~M$_\odot$, see \S\ref{Mass Distribution}) also agrees with this within its uncertainty of a factor of three, indicating that the entire gas sampled by C$^{18}$O represents the HMC SiO-NE.
We note that, compared with low mass star forming molecular cores, which contain $\sim3-30$~M$_\odot$ inside the radius of $\sim0.1$~pc, some 1000 times mass is concentrated in the similar radius in the case of massive star forming cores.

\subsection{Position velocity diagrams}\label{Results-PV}

\begin{figure*}[htbp]
\includegraphics*[bb= 00 430 500 710, scale=0.8]{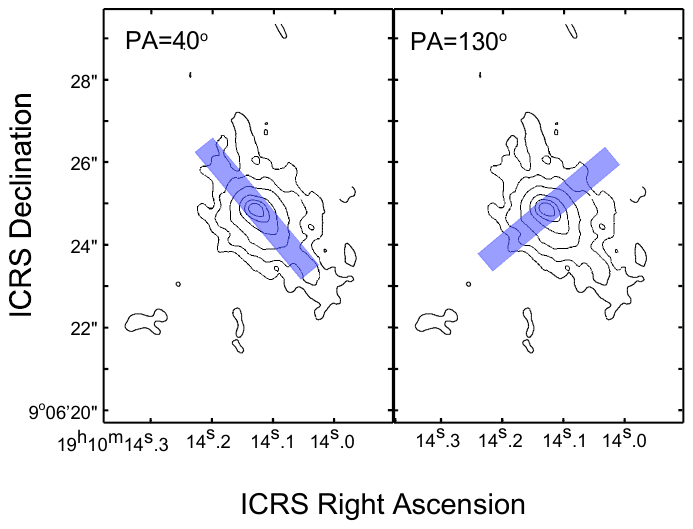}
\caption{Two strips for the position velocity diagrams shown in Fig~\ref{Fig11} (PA=40\degree) and Fig~\ref{Fig12} (PA=130\degree) superposed on the contours (5, 10, 20, 40, 60, and 80\% levels) of the continuum emission.
The slit length is 4\farcs0 and the averaging slit width is 0\farcs5.
\label{Fig10}}
\end{figure*}

\begin{figure*}[htbp]
\includegraphics*[bb= 50 100 700 690, scale=0.8]{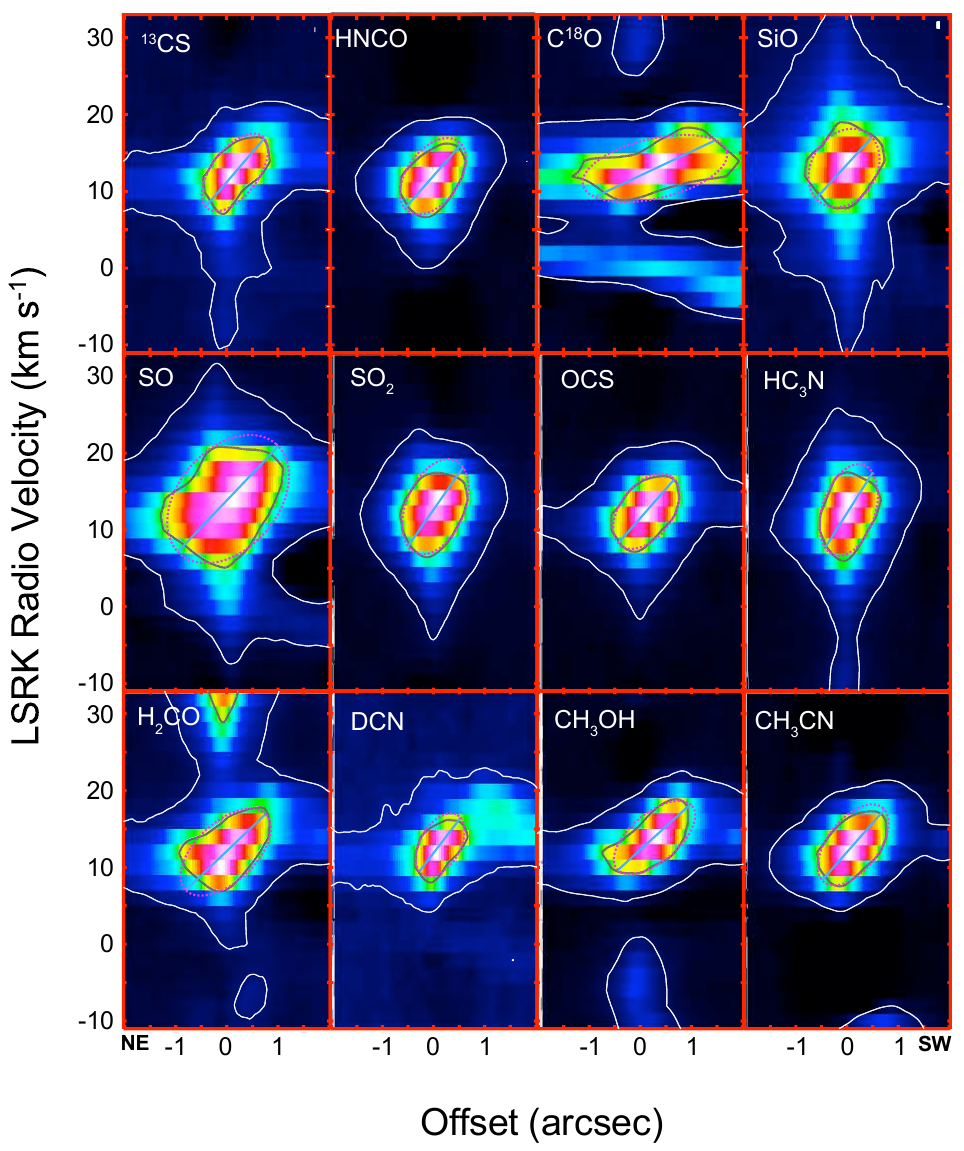}
\caption{Position velocity diagrams along the major axis (PA=40\degree) of the flattened envelope for the molecular line emissions noted on each panel.
The gray and white contours show the 50\% and 5\% level, respectively, of the peak brightness.
The red dotted ellipse on each panel is the 50\% level contour of the two dimensional gaussian fitted to the PV data, with the blue straight line representing the ridge line, from which the effective radius and radial velocity are derived.
Positions with negative offsets are the NE side.
The CH$_{3}$CN diagram is for $K$=3.
For the H$_{2}$CO panel, the emission feature at $V_{\rm LSR}\,\ga\,25$\,km\,s$^{-1}$ does not belong to H$_{2}$CO.
\label{Fig11}}
\end{figure*}

We use position-velocity (PV) diagrams along the two strips shown in Figure~\ref{Fig10} in order to examine the rotation of the flattened envelope in more detail. 
Figure~\ref{Fig11} shows the PV diagrams along the major axis of the dust emission (PA=40\degree), with the 50\% (gray) and 5\% (white) level contours explicitly drawn on each panel.
As expected, the velocity gradient along the major axis is obvious from the tilted  ridges of PV emission.
The magnitude of each velocity gradient is different, reflecting that each molecular line samples a specific radial region of the flattened envelope rotating at a different velocity.
We will analyze this in the next subsection \S\ref{Results-Rotation}.

Other than the prominent tilted ridges of PV emission, we see two types of faint features delineated by the 5\% level contours in Figure~\ref{Fig11}.
First, many lines show faint, high velocity emission at $V_{\rm LSR}$\,\ga\,22\,km\,s$^{-1}$ and/or $V_{\rm LSR}$\,\la\,2\,km\,s$^{-1}$ toward the continuum peak, namely, faint features extending along the vertical axis at the positional offset of 0$''$.
For the compact emission molecules, such high velocity emission  may arise from the rapidly rotating inner part ($R\,\la\,1,000$~au) of the envelope.
For SiO and SO, the high velocity emission comes from the outflow at its accelerating region as is evident from the channel maps (Figures~\ref{Fig07} and \ref{Fig08}).

\begin{figure*}[htbp]
\includegraphics*[bb= 50 100 700 690, scale=0.8]{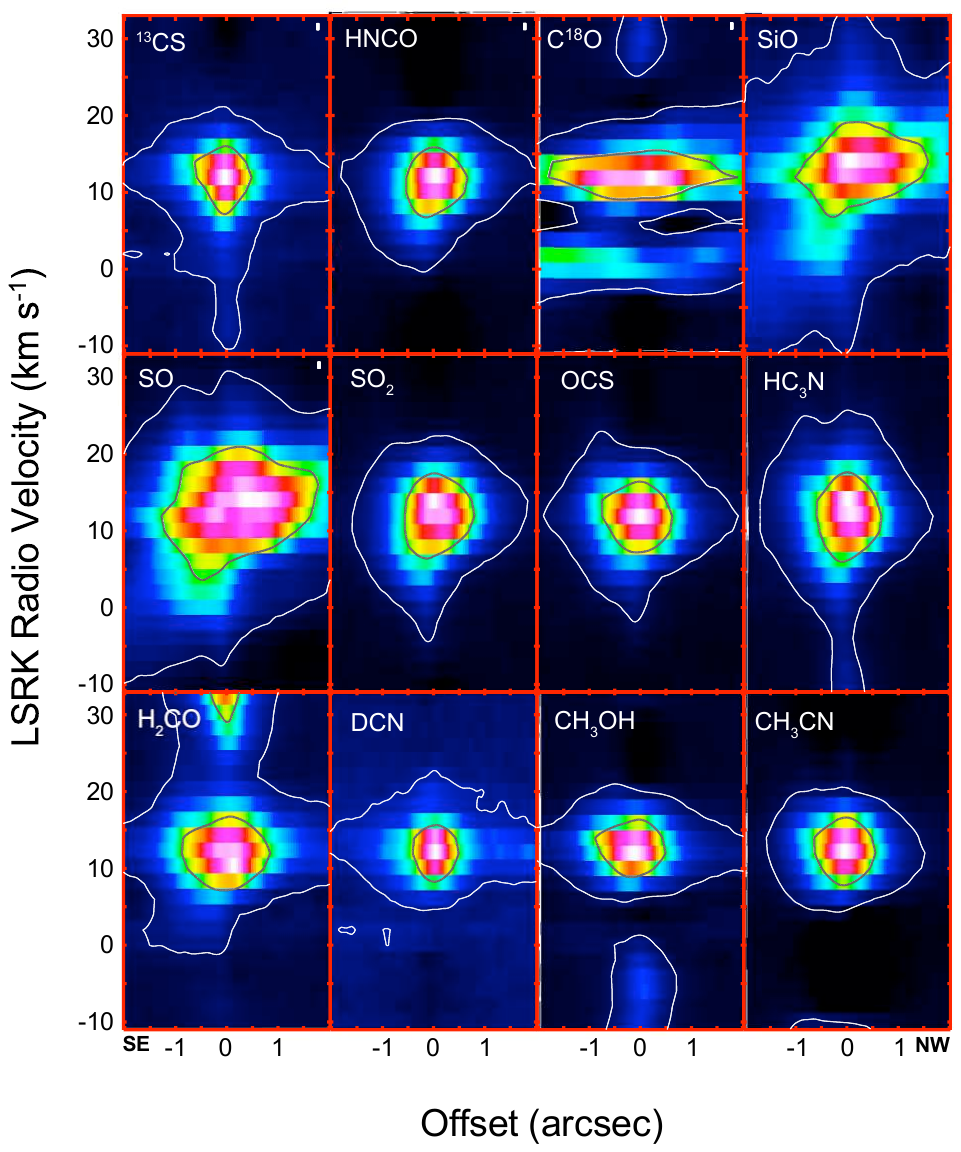}
\caption{Position velocity diagrams along the minor axis (PA=130\degree) of the flattened envelope for the molecular line emissions noted on each panel.
The gray and white contours show the 50\% and 5\% level, respectively, of the peak brightness for each panel.
Positions with negative offsets are the SE side.
The CH$_{3}$CN diagram is for $K$=3.
For the H$_{2}$CO panel, the emission feature at $V_{\rm LSR}\,\ga\,25$\,km\,s$^{-1}$ does not belong to H$_{2}$CO.
\label{Fig12}}
\end{figure*}

Second, at the absolute offset larger than 1$''$ corresponding to $R$\,\ga\,11,000~au, the lines of $^{13}$CS, C$^{18}$O, SiO, SO, OCS, H$_{2}$CO, DCN, and CH$_{3}$OH are associated with faint, horizontal features around the systemic velocity of $\sim$12\,km\,s$^{-1}$.
The faint emission has a radial velocity of $V_{\rm LSR}\sim$11\,km\,s$^{-1}$ on the NE side (negative offsets) and $\sim$14\,km\,s$^{-1}$ on the SW side.
Such a velocity difference at outer radii is not evident along the minor axis (Figure~\ref{Fig12}) for the compact emission molecules.
Thus the velocity shift is due to rotation in the outer part of MCN-a.
The inclination-corrected rotation velocity is then $\sim$2\,km\,s$^{-1}$ there.
This means that the outer part ($R\,\ga\,11,000$~au) of the flattened envelope is rotating more slowly than its inside.
The two faint features of the PV diagrams hence show an overall trend that the rotation velocity increases from its outer part to the center.
The tilted ridge emissions, however, reveal a different rotation law from this tendency as we will see in the next subsection \S\ref{Results-Rotation}.

Figure~\ref{Fig12} shows the position velocity diagrams along the minor axis of the dust emission (PA=130\degree).
We do not see any clear velocity gradient for the lines of compact emission molecules.
Consequently, the flattened envelope shows no detectable radial motion with the current velocity resolution of 2\,km\,s$^{-1}$.
This means that rotational motion is dominant in the envelope, but does not exclude the possibility that it undergoes accretion at a sufficiently high rate to feed the massive central protostar.
We will return to this point in \S\ref{Disk accretion}.

For SiO, we see an overall velocity gradient as a result of the outflow, namely the SE side (negative offsets) tends to be blueshifted and the NW side tend to be redshifted.
Blueshifted emission features extending to the SE are seen both in SiO and SO.
The outflow interpretation is also supported by the apparent ``acceleration,'' particularly visible in the PV diagram of SiO, often observed for molecular outflows that are indirectly accelerated by higher velocity winds nested inside them \citep[e.g.,][]{Lada1996}.

\subsection{Rotation of the flattened envelope}\label{Results-Rotation}

For each molecular transition presented in Figure~\ref{Fig11}, we measured an effective radius ($R$) and a radial velocity ($V_{\rm rot}\sin i$) at the radius.
By fitting an elliptical gaussian to each PV diagram, we defined a ridge line along the velocity gradient of PV emission.
Methodological details of determining the ridge lines is provided in Appendix~\ref{Ridge line}.
We superposed on Figure~\ref{Fig11} blue straight lines representing the ridge lines together with the red dotted ellipses representing the 50\% level contours of two dimensional gaussians fitted to the PV data.

For each transition, we read the positional offsets of the ridge line on its positive and negative sides defined by the 50\% contour of the fitted ellipse, thus measuring the FWHM spatial extent of each emission.
We then set the effective radius of rotation as one half of the FWHM extent.
The overall uncertainty of the measured radii was $\sim$10\%, inferred from the difference of contours between the observed and fitted data.
Similarly, we measured the FWHM extent of the ridge line along the radial velocity axis and set $V_{\rm rot}\sin i$ as its one half.
The uncertainties in radius and velocity measurements are thus correlated.
The radius, velocity, and the inclination-corrected rotation velocity are listed in Table~\ref{Table02} as Rotation Radius, $V_{\rm rot}\sin i$, and $V_{\rm rot}$.

\begin{table*}[ht]
\begin{minipage}{\textwidth}
\caption{Properties of the rotating envelope derived from molecular lines}
\label{Table02}
\begin{center}
\scalebox{0.86}[0.86]{
\begin{tabular}{llcccccccl}

\hline\hline
Molecule & Transition & Peak & Integrated&Minor to  & Rotation & $V_{\rm rot}\sin i$  & $V_{\rm rot}$ &  Dynamical  \\
               &          & Brightness    &   Intensity  &Major &     Radius &          &           & Mass   \\
  &  & (K) & (K km\,s$^{-1}$) & Axial Ratio &  (au) &  (km\,s$^{-1}$) &  (km\,s$^{-1}$) &  (M$_\odot$) \\
\hline 
$^{13}$CS & $J=5-4$ & 67 & 641& 0.669& 6,470 & 4.38 & 5.95 & 257 \\
HNCO      & $J_{K_a,K_c}=10_{0,10}-9_{0,9}$ & 103 & 893 &0.898&  6,290 	& 4.20 & 5.70 & 230  \\
C$^{18}$O & $J=2-1$ &82  & 531 &  --\footnotemark[*]  & 13,400 & 3.81 & 5.16 & 401 \\
SiO       & $J=5-4$ &60  & 865  &--\footnotemark[*]  & 6,090 & 4.02 & 5.45 & 204 \footnotemark[$\dagger$]  \\
SO        & $N_J=6_5-5_4$ &  138& 1180 &  --\footnotemark[*] & 11,120 & 6.96 & 9.43 & 1115 \\
SO$_2$    & $J_{K_a,K_c}=28_{3,25}-28_{2,26}$ & 115 &1,266  &0.898  & 6,120 & 5.19 & 7.04 & 342 \\
OCS       & $J=19-18$ &139 & 1,349 &0.877 & 6,250 & 4.23 & 5.74 & 232 \\
HC$_3$N   & $J=24-23$ & 119 &1,452  &0.761 & 5,240 & 4.96 & 6.73 & 267 \\
H$_2$CO   & $J_K=3_{22}-2_{21}$ & 59 & 613 & 0.790& 8,830 & 5.03 & 6.83 & 463 \\
DCN       & $J=5-4$ &31  & 235 &  0.604  & 4,570 & 3.55 & 4.82 & 120 \\
CH$_3$OH  & $J_K=4_3-3_1$ & 99 & 687 &0.723 & 8,110 & 4.34 & 5.89 & 317 \\
CH$_3$CN  & $J_K=12_3-11_3$ & 133 & 898  & 0.906 & 6,770 & 4.55 & 6.18 & 291 \\
CH$_3$CN  & $J_K=12_4-11_4$ & 103 &  652 & 0.965  & 6,090 & 4.15 & 5.63 & 217 \\
CH$_3$CN  & $J_K=12_7-11_7$ & 33 & 247 & 0.863 & 3,390 & 2.78 & 3.78 & 54.5 \\
CH$_3$CN  & $J_K=12_8-11_8$ &  14  &  88  &0.891 & 2,440 & 2.38 & 3.23 & 28.6  \\
\hline
\end{tabular}
}
\end{center}
\footnotetext{}{$^{*}$Outflow is dominant. \\
$^{\dagger}$Value should be taken as nominal because the  rotation of the envelope is not evident in the channel map. }\\

\end{minipage}
\end{table*}

\begin{figure}[htbp]
\includegraphics[bb= 50 20 600 520, scale=0.5]{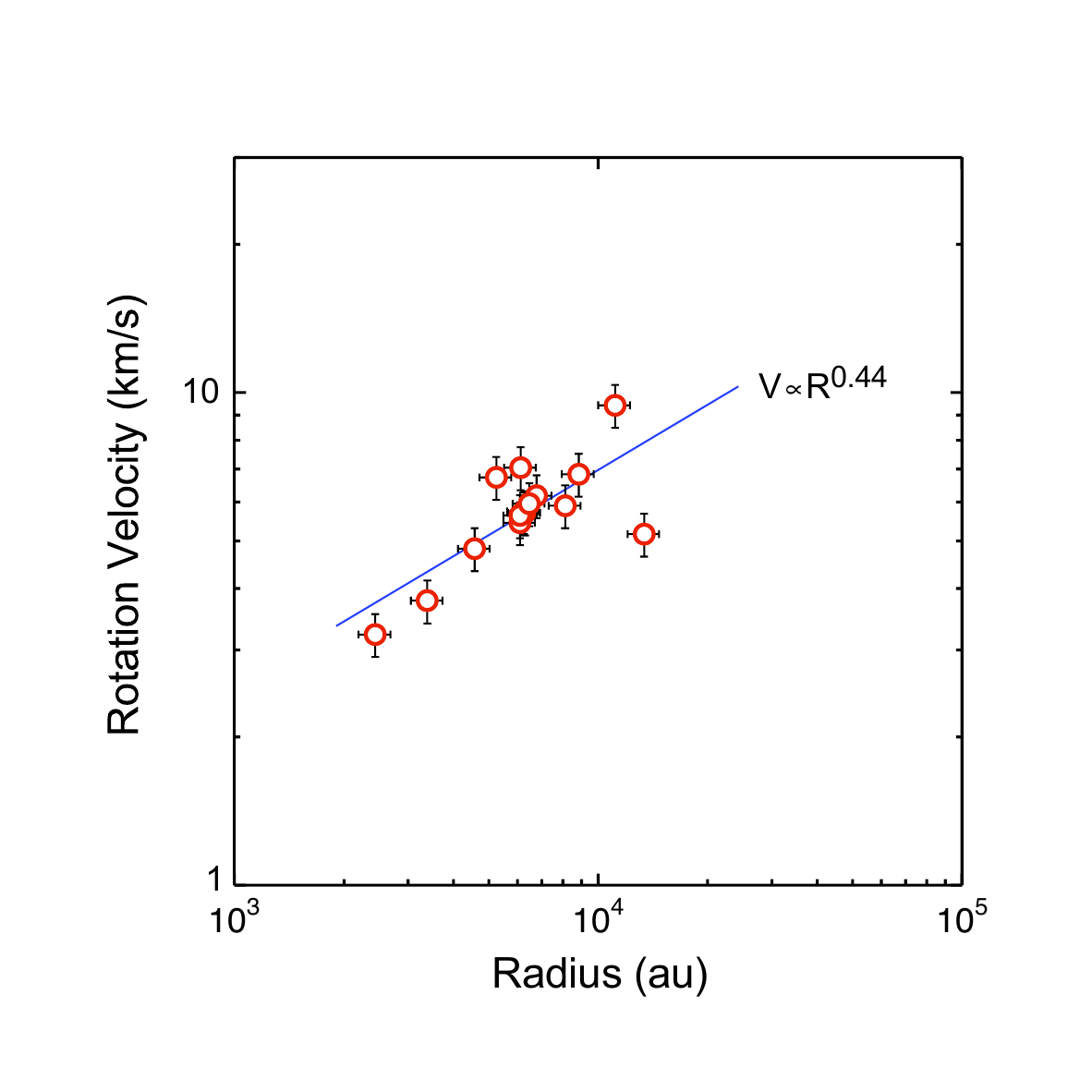}
\caption{Rotation velocity as a function of radius.
The blue curve shows a power law fit.
}
\label{Fig13}
\end{figure}

Figure~\ref{Fig13} shows the rotation velocity as a function of radius.
We obtain a rotation curve as $V_{\rm rot}\propto R^{0.44\pm0.11}$ for 2,400~au \la\,$R$\,\la\ 14,000~au by assuming a power law relation between the two variables. 
If we give the coefficient for the specific power of 0.44, the rotation law becomes 
\begin{equation}\label{eq:rotation curve}
V_{\rm rot}=4.11^{+0.80}_{-0.67}\,(R\,[{\rm au}]/3,000)^{0.44}$ km\,s$^{-1}.
\end{equation}
In \S\ref{Mass Distribution}, we will derive the dynamical mass from the rotation curve and compare it with the mass estimated from the dust and line emissions.

The derived rotation velocity monotonically increases as $R$ increases from 2,400~au to 14,000~au.
This seems to be opposite to the overall trend that the rotation velocity increases from the outer part to the center as we saw in \S\ref{Results-PV}.
In fact, the above derived rotation law must break up at both larger and smaller radii.
As we will discuss in \S\ref{Mass Distribution}, the break-up at a smaller radius occurs around $R\sim$ 1,000~au, where the flattened envelope turns into a non self-gravitating Keplerian-like rotating disk.
The break-up at a larger radius should be located around $R\sim$ 11,000~au (Offset$\sim 1''$), outside of which the faint emission from slowly rotating gas becomes discernible in the PV diagrams.
We speculate that freely infalling ambient gas begins to settle in the rotation-dominated, self-gravitating envelope at this radius.
The opposite trend of the observed rotation velocity with radius could consequently be due to the pile-up of mass in the intermediate range of radius.

We should note that the outermost data point for C$^{18}$O in Figure~\ref{Fig13} is deviated relatively largely from the trend defined by the other data points.
Its radius is $R=13,400$~au, and the rotation velocity there may be outside the range that the above monotonically increasing rotation law applies.
If we ignore this data point, we get a different power law dependence as $V_{\rm rot}\propto R^{0.63\pm0.08}$.
The dynamical mass we derive in \S\ref{Mass Distribution} as well as the results we obtain there are basically the same for the two power law indices of 0.44 and 0.63.
We thus include the C$^{18}$O data point for our analysis in order to avoid arbitrariness in handling data and use the power law $V_{\rm rot}\propto R^{0.44}$ hereafter.

\subsection{Rotation temperature}\label{Trot}

Many rotational transitions of CH$_3$CN, HCNO, and CH$_3$OH were detected in the total band width of the current data.
Because these lines originate from different energy levels, we can use their intensities to estimate the rotation temperature of each molecular species.
We used the rotation diagram analysis introduced by \citet{Hollis1982} and subsequently developed by a number of authors \citep[e.g.][]{Loren1984, Turner1991, Goldsmith1999}.

We calculated the upper state column density $N_{\rm u}$ of each transition from the total integrated intensity $\int_{-\infty}^{\infty}T_{\rm b}\,dv$ (in unit of K\,km~s$^{-1}$) within the beam solid angle ($\Omega_{\rm beam}$) toward the peak using the following formula for optically thin emission, 
\begin{equation}\label{eq:column_density}
N_{\rm u} = \frac{8\pi\,k\,\nu_{\rm ul}^2}{hc^3A_{\rm ul}}\int_{-\infty}^{\infty}T_{\rm b}\,dv,
\end{equation}
where $\nu_{\rm ul}$ is the frequency of the transition and $A_{\rm ul}$ is the Einstein A coefficient. 
The column density $N_{\rm u}$ is related to the rotation temperature $T_{\rm rot}$ as,
\begin{equation}\label{eq:boltzmann_relation}
N_{\rm u}=\frac{N g_{\rm u}}{Q(T_{\rm rot})} e^{-E_{\rm u}/k\,T_{\rm rot}},
\end{equation}
where $N$ is the column density of the molecular species, $g_{\rm u}$ is the degeneracy of the upper state, $E_{\rm u}$ is the upper state energy, and $Q(T_{\rm rot})$ is the partition function. 
Assuming that $T_{\rm rot}$ is the same for all the observed transitions of the molecular species, we obtain $T_{\rm rot}$ from the relation between $\ln(N_{\rm u})$ and $E_{\rm u}$ by applying a straight line fit to the plot of $\ln(N_{\rm u}/g_{\rm u})$ versus $E_{\rm u}/k$, which has a slope of $1/T_{\rm rot}$ and an intercept of $\ln(N/Q(T_{\rm rot}))$. 
The data we used are summarized in Table~\ref{Table03}. 
The optically thin assumption may be justified, at least for CH$_3$CN, from the observed CH$_3$CN to CH$_3$$^{13}$CN intensity ratios of $\sim$40.
In addition, we do not see in our data that the rotation temperatures derived for the low $K$ components are systematically larger than those of the high $K$ components, which is expected to occur if high optical depths are affecting the temperature estimates \citep{Araya2005}.
The optical depth effects are thus not significant.

\begin{table*}[ht]
\renewcommand{\arraystretch}{0.7}
\begin{minipage}{\textwidth}
\caption{Data for rotation energy level population analysis}
\label{Table03}
\begin{center}
\scalebox{0.9}[0.9]{

\begin{tabular}{llccccccl}

\hline\hline

Molecule & Transition & Frequency & Detection & $E_{\rm u}$  & $\int T_{\rm b} dV$  & ln($N_{\rm u}/g_{\rm u}$) & Remarks \\
               &                 &  (GHz)        &                  &(K)           &   (K~km\, s$^{-1}$)   &  & \\
\hline
CH$_3$CN &  $J_K=12_0-11_0$& 220.7472612 & Yes & 68.8664  & 1080 &  31.41 & blended \\
CH$_3$CN &  $J_K=12_1-11_1$& 220.7430106 & Yes & 76.0111  & 954 &  31.30 &  blended\\
CH$_3$CN &  $J_K=12_2-11_2$ & 220.7302607 & Yes & 97.4433  & 687 &  30.99 &  \\
CH$_3$CN &  $J_K=12_3-11_3$ & 220.7090165 & Yes & 133.1586  & 897 &  31.29 &  \\
CH$_3$CN &  $J_K=12_4-11_4$ & 220.6792869 & Yes & 183.1483  & 652 & 31.03 &  \\
CH$_3$CN &  $J_K=12_5-11_5$& 220.6410839 &Yes  & 247.4016  & 504 &  30.84 & blended \\
CH$_3$CN &  $J_K=12_6-11_6$ & 220.5944231 & ?  & 325.9034  & -- &--  & blended \\
CH$_3$CN &  $J_K=12_7-11_7$& 220.5393235 & Yes & 418.6361  & 247  & 30.36 &  \\
CH$_3$CN & $J_K=12_8-11_8$ & 220.4758072 &Yes  & 525.5787  & 88 & 29.50 &  \\
CH$_3$CN &  $J_K=12_9-11_9$ & 220.4039000 & ? & 646.7066  & -- & -- & blended \\
CH$_3$CN &  $J_K=12_{10}-11_{10}$ & 220.3236306 & No & 781.9927  & -- & -- &  \\
CH$_3$CN & $J_K=12_{11}-11_{11}$ & 220.2350310 & No & 931.4061  & -- & --  &  \\
\hline
$T_{\rm rot}$(CH$_3$CN) &  &  &  &  &  &  &  & 278~K \\
 \hline
HNCO &  $J_{K_a,K_c}=10_{0,10}-9_{0,9}$ & 219.7982740 & Yes & 58.0192  & 893 &31.41 &  \\
HNCO &  $J_{K_a,K_c}=10_{1,10}-9_{1,9}$ & 218.9810090 &Yes  & 101.0788  &878 & 31.41 &  \\
HNCO & $J_{K_a,K_c}=10_{1,9}-9_{1,8}$ & 220.5847510 & Yes & 101.5022  & 807 &  31.32 &  \\
HNCO & $J_{K_a,K_c}=10_{2,9}-9_{2,8}$ & 219.7338500 & Yes & 228.2847  & 702 &  31.21 &  \\
HNCO & $J_{K_a,K_c}=10_{2,8}-9_{2,7}$ & 219.7371930 & Yes & 228.2851  & 287 & 30.32 & blended \\
HNCO & $J_{K_a,K_c}=10_{3,8}-9_{3,7}$ & 219.6567695 & Yes & 432.9598  & 279 &  30.34 &  blended\\
HNCO & $J_{K_a,K_c}=10_{3,7}-9_{3,6}$ & 219.6567708 & Yes & 432.9598  & 297 &  30.40 & blended \\
HNCO & $J_{K_a,K_c}=10_{4,6}-9_{4,5}$ & 219.5470820 & Yes & 708.7094  & 77 &  29.13 &  \\
HNCO & $J_{K_a,K_c}=10_{5,5}-9_{5,4}$ & 219.3924120 & No & 1049.5365  &  -- &    -- &  \\
HNCO & $J_{K_a,K_c}=10_{6,4}-9_{6,3}$ & 219.1326788 & No & 1450.3262  &  -- &   -- &  \\
\hline
$T_{\rm rot}$(HNCO)&  &  &  &  &  &  &  & 297~K \\
 \hline
CH$_3$OH & $J_K=4_3-3_1$ E1 vt=0 & 218.4400630 &Yes  & 45.4599  & 687 & 32.36 &  \\
CH$_3$OH & $J_K=8_0-7_1$ E1 vt=0 & 220.0785610 & Yes & 96.6133  & 618 &  31.26 &  \\
CH$_3$OH & $J_K=10_{2-}-9_{3-}$ vt=0 & 231.2811100 & Yes & 165.3471  & 561 &  30.94 &  \\
CH$_3$OH & $J_K=10_{2+}-9_{3+}$ vt=0 & 232.4185210 & Yes & 165.4017  & 393 &  30.58 &  \\
CH$_3$OH & $J_K=10_{3+}-11_{1+}$ vt=0 & 230.9612740 & Yes & 177.4534  & 55 &  28.67 &  \\
CH$_3$OH & $J_K=10_{-3}-11_{-2}$ E2 vt=0 & 232.9457970 & Yes & 190.3695  & 393 &  30.63 &  \\
CH$_3$OH & $J_K=11_{3}-10_{-3}$ E2 vt=0 & 219.8524230 & No & 200.9203  &   -- &      -- &  \\
CH$_3$OH & $J_K=10_{-5}-11_{-4}$ E2 vt=0 & 220.4013170 & ? & 251.6432  &   -- &      -- & blended \\
CH$_3$OH & $J_K=16_{1}-15_{3}$ E1 vt=0 & 217.5250020 & No & 336.7162  &   -- &     -- &  \\
CH$_3$OH & $J_K=16_{2}-16_{0}$ E1 vt=0 & 219.0927420 & No & 338.1382  &    --&    -- &  \\
CH$_3$OH & $J_K=18_{2}-18_{0}$ E1 vt=0 & 232.8547910 & No & 419.3985  &   -- &   --  &  \\
CH$_3$OH & $J_K=18_{3-}-17_{4-}$ vt=0 & 233.7956660 & Yes & 446.5802  & 103 &  28.63 &  \\
CH$_3$OH & $J_K=20_{1}-20_{0}$ E1 vt=0 & 217.8865040 & Yes & 508.3758  & 155 & 28.98 &  \\
CH$_3$OH & $J_K=21_{4-}-22_{1-}$ vt=0 & 219.9163990 & No & 616.2585  &  --  &     -- &  \\
CH$_3$OH & $J_K=23_{2+}-22_{4+}$ vt=0 & 231.7625490 & No & 678.4020  &   -- &    --&  \\
\hline
$T_{\rm rot}$(CH$_3$OH)&  &  &  &  &  &  &  & 154~K \\
\hline
\end{tabular}
}
\end{center}
\end{minipage}

\end{table*}

\begin{figure}[htbp]
\includegraphics[bb= 50 170 300 580, scale=0.45]{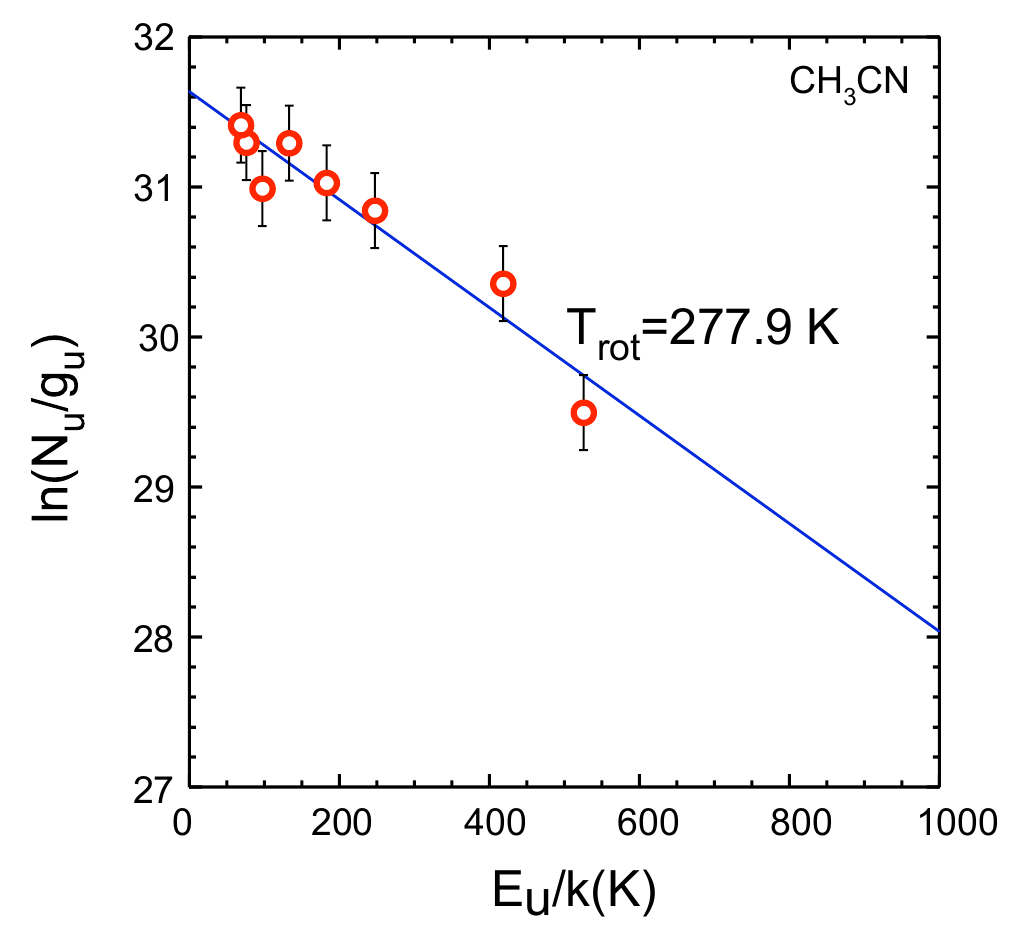}
\caption{Rotation energy level population diagram for CH$_3$CN.
}
\label{Fig14} 
\end{figure}

\begin{figure}[htbp]
\includegraphics[bb= 50 170 300 580, scale=0.45]{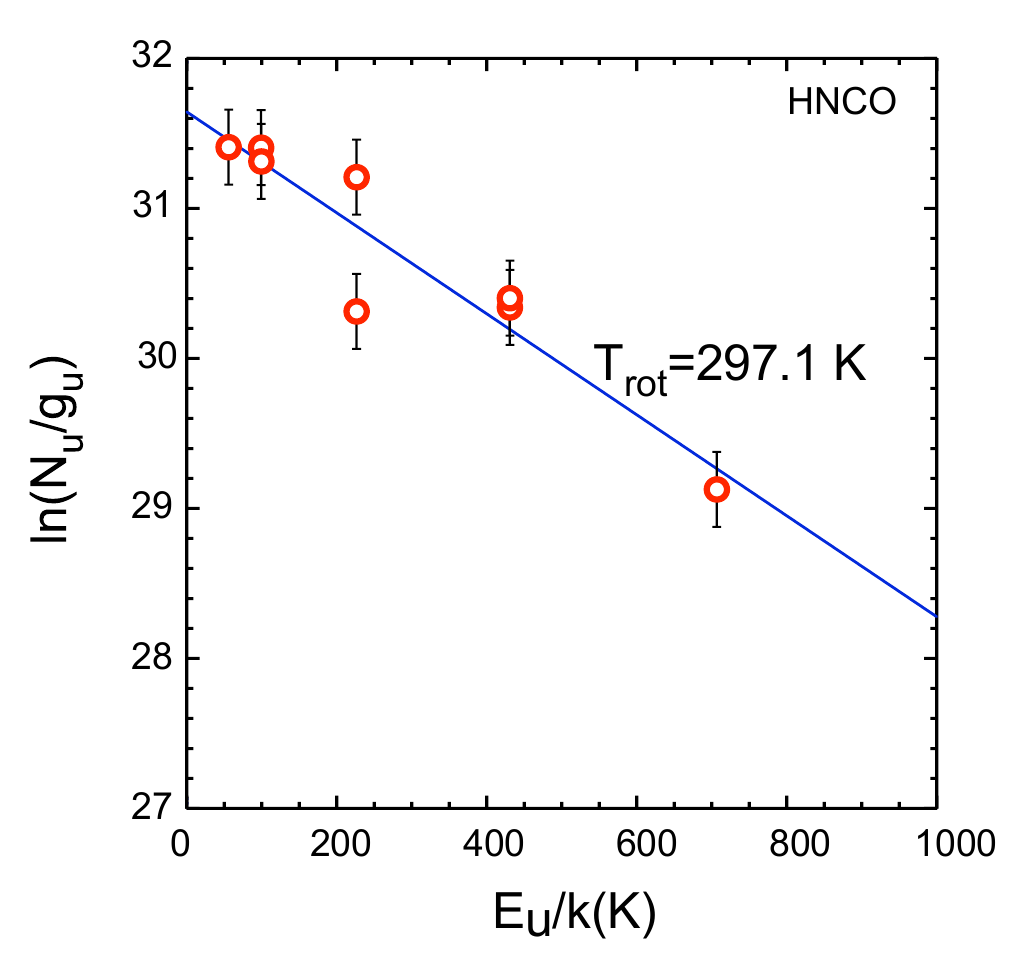}
\caption{Rotation energy level population diagram for HNCO.
}
\label{Fig15}
\end{figure}

\begin{figure}[htbp]
\includegraphics[bb=50 170 300 580, scale=0.45]{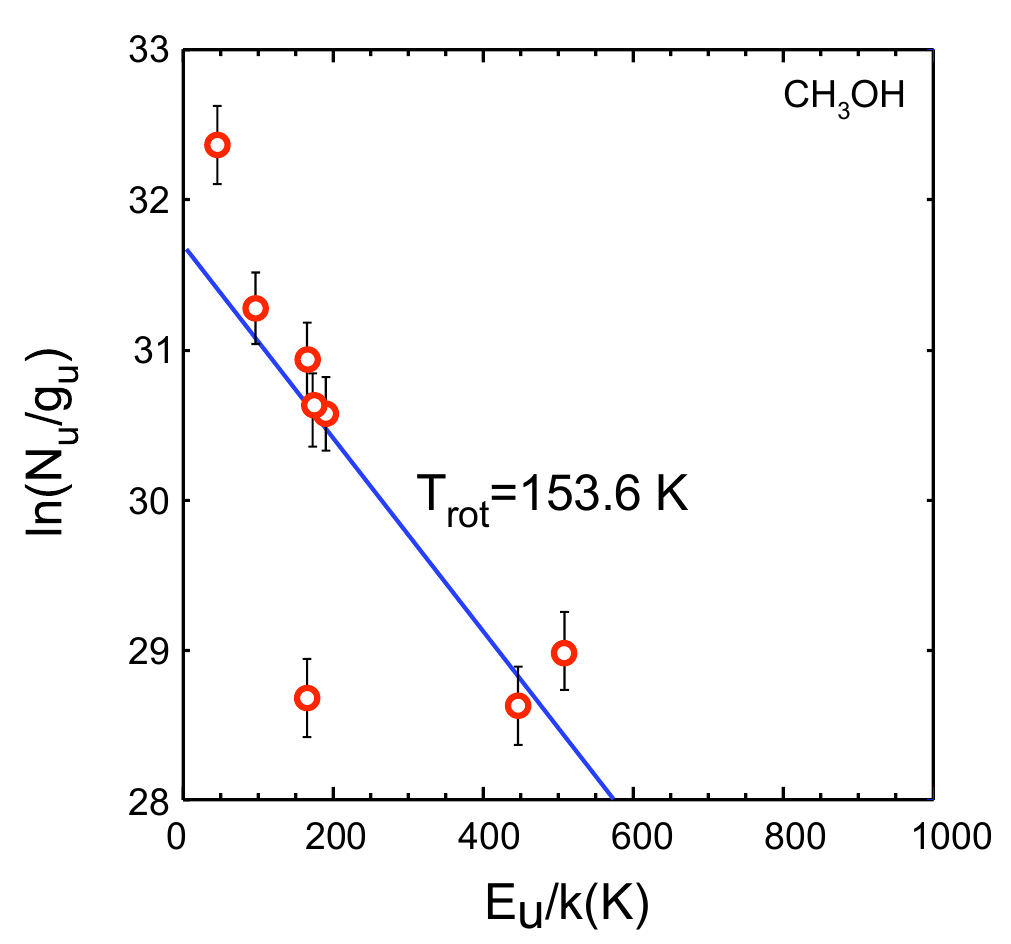}
\caption{Rotation energy level population diagram for CH$_3$OH.
}
\label{Fig16}
\end{figure}

Rotation energy level population diagrams for CH$_3$CN, HNCO, and CH$_3$OH are presented in Figures~\ref{Fig14}, \ref{Fig15} and \ref{Fig16}, respectively.
We obtained $T_{\rm rot}$(CH$_3$CN) = 278\,$^{+39}_{-30}$~K, $T_{\rm rot}$(HNCO) = 297\,$^{+52}_{-39}$~K, and $T_{\rm rot}$(CH$_3$OH) = 154\,$^{+73}_{-37}$~K.
With the assumption of LTE, these values should be regarded as representing the temperature at the radius $R\,\la\,$ 1,700~au of the beam solid angle.

We compare the rotation temperatures with the dust temperature distribution.
We first consider the case where the dust particles are under radiative equilibrium with a star whose effective temperature and radius are $T_*$ and $R_*$, respectively.
In this case, the envelope is assumed to be optically thin, so that the dust particles are directly heated by the star.
Although such a case is unrealistic for a heavily embedded protostar, we would like to show that this is the case where we can reconcile the apparent central stellar luminosity ($\sim3\times10^4$~L$_\odot$) with the rotation temperatures we observed at $R\,\sim$\,1,700~au.

In this case, the temperature of dust particles with the emissivity spectral index $\beta$ located at a distance $R$ from the star is approximated by 
\begin{equation}\label{eq:dust temperature}
T_{\rm dust}(R)=\Big(\frac{R_*}{2R}\Big)^{2/(4+\beta)}\,T_{*},
\end{equation}
where $(R_*/2R)^2$ is the dilution factor \citep[e.g.,][]{Beckwith1990, Spitzer1978} and $\beta$ is assumed to be 2 (see \S\ref{Disk Mass}).
Assuming that at least a B0 star is embedded at the center of MCN-a, which coincides with the UCHII region J$_{1}$ \citep{DePree1997, Miyawaki2022}, we may use $T_*=27,000$~K and $R_*=7$~R$_\odot$ ($L_{\rm bol}=23,000$\,L$_\odot$) \citep[e.g.,][]{Hohle2010} to estimate the dust temperature.
We then obtain $T_{\rm dust} = 570$~K at the envelope radius  $R$\,=\,1,700~au.

It may be better to examine one more case for the central star, because the dust temperature at a given distance from the star depends not only on its bolometric luminosity, but also on its surface temperature (or radius), reflecting the non-blackbody optical properties of dust particles.
We should pick up one more case with different surface temperature, but ideally with the same bolometric luminosity.
We choose an accreting protostar with $T_*=7,500$~K and $R_*=100$~R$_\odot$ ($L_{\rm bol}=28,000$\,L$_\odot$) \citep{Hosokawa2009} as the lowest surface temperature case so all the other cases lie between the two models considered here.
The dust temperature in this case is $T_{\rm dust}\,=\,390$~K at $R$\,=\,1,700~au.
We should note that a star with such a low surface temperature cannot produce a UCHII region observed toward MCN-a.
 
The rotation temperature of $T_{\rm rot}\sim300$~K at $R\,\sim$\,1,700~au measured with CH$_3$CN and HNCO lines agrees with the embedded protostar model of $T_{\rm dust}=390$~K within the errors.
The rotation temperature of $T_{\rm rot}\sim160$~K measured with the CH$_3$OH lines is somewhat smaller than the embedded protostar model.
The embedded B0 star model gives a higher dust temperature: $T_{\rm dust}$\,=\,570~K compared with $T_{\rm rot}\sim300$~K at $R\,\sim$\,1,700~au.

When we consider a more realistic temperature distribution of an envelope around a heavily embedded massive star, we need a luminosity much higher than that of a B0 star in order to obtain the dust temperature of 300~K at 1,700~au.
Radiative transfer calculations show that a O6 ZAMS star of $\sim2.5\times10^5$~L$_\odot$ barely heats the dust particles to $\sim$300~K at $R\,\sim$\,1,700~au \citep[e.g.,][]{Indebetouw2006}. 
In addition, far-infrared observations with {\it SOFIA} suggested that the extinction-corrected bolometric luminosity of MCN-a could be as large as 2$\times 10^{6}$~L$_\odot$ \citep{DeBuizer2021}.

\begin{figure}[htbp]
\includegraphics[bb=50 30 300 530, scale=0.4]{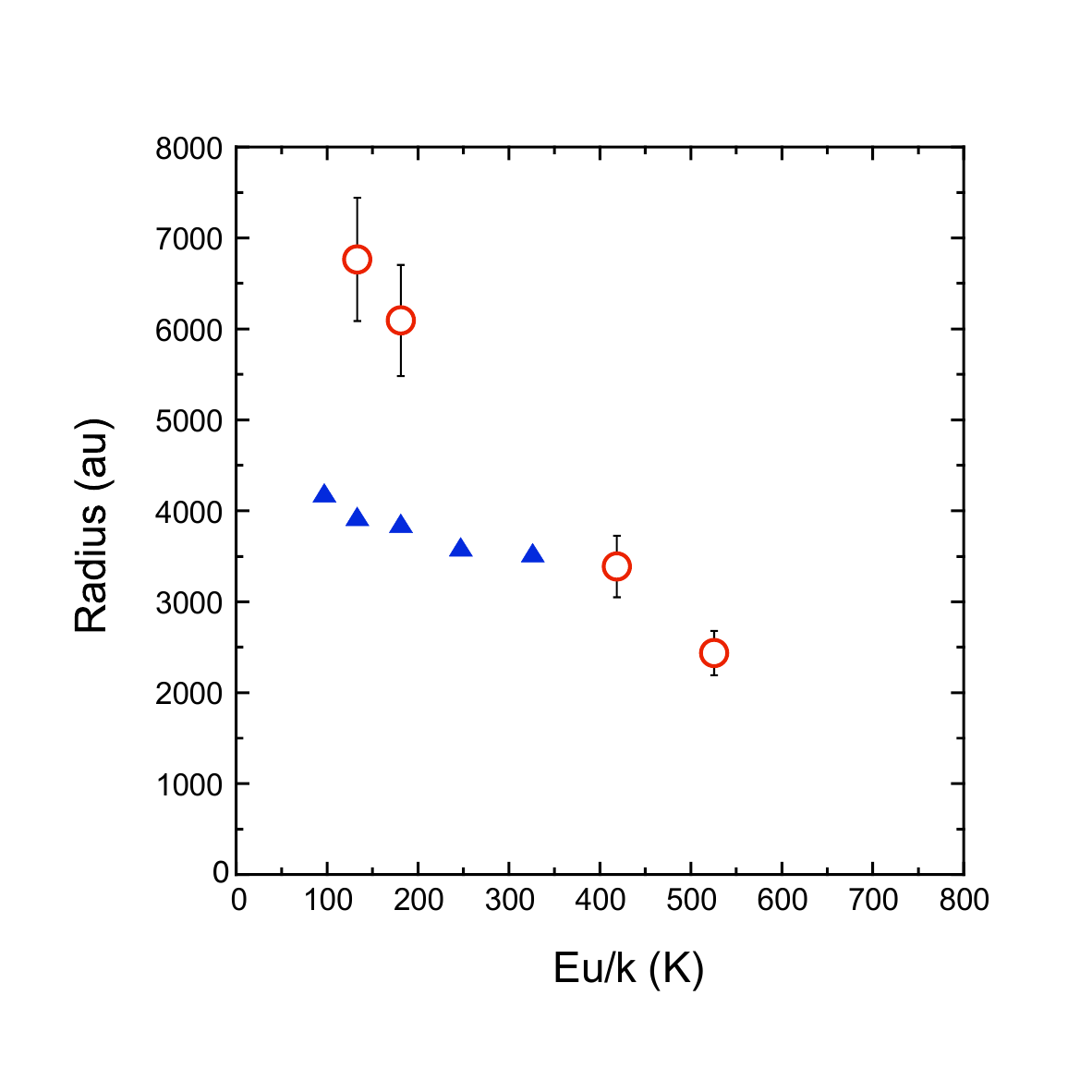}
\caption{Rotation radius is plotted against the upper state energy for the four CH$_{3}$CN ($J=12-11$) transitions of $K$=3, 4, 7, and 8.
Blue triangles show a similar plot for the CH$_{3}$CN ($J=12-11$) transitions of  $K$=2, 3, 4, 5 and 6 observed toward G24.78+0.08 A1 \citep{Moscadelli2021}.
}
\label{Fig17}
\end{figure}

The high luminosity of the (proto)star at the center of MCN-a
is also supported when we compare the current data with previous  
observations.
Figure~\ref{Fig17} shows a plot of the rotation radius (taken from Table~\ref{Table02}) against the upper state energy $E_{\rm u}$ (taken from Table~\ref{Table03}) for the four CH$_{3}$CN ($J=12-11$) transitions of $K$=3, 4, 7, and 8.
It shows a trend that the emission radius decreases from $R\sim$~7,000~au  to 2,400~au as the upper state energy increases from $E_{\rm u}/k =$~100~K to 500~K. 
Similar trends were previously reported by various authors \citep{Cesaroni2014, Moscadelli2021} and are interpreted as reflecting the decrease of gas temperature with radius.

According to \citet{Moscadelli2021}, who reported a similar tendency using the same CH$_{3}$CN ($J=12-11$) transitions as ours toward the HCHII region G24.78+0.08 A1 as also plotted  in Figure\ref{Fig17}, the radial extent of emission decreases from $\sim$4,200~au to $\sim$3,600~au as $E_{\rm u}/k$ varies from 100~K to 300~K.
Their relation gives a significantly smaller radius at a given $E_{\rm u}/k$ than does the MCN-a case.
The larger radius of MCN-a for a given upper state energy indicates that its bolometric luminosity should be larger than that of G24.78+0.08 A1, which carries much of the total bolometric luminosity of $2\times 10^5$~L$_\odot$ for G24.78+0.08 \citep{Moscadelli2021}.
This suggests that the true bolometric luminosity of MCN-a is more than 10 times larger than its apparent luminosity of $\sim3\times 10^{4}$~L$_\odot$ \citep{DeBuizer2021}.
These pieces of evidence support that the central source of MCN-a is a very luminous massive young (proto)star still deeply embedded in a HMC.

\section{Discussion}\label{DISCUSSION}


\subsection{Mass of the flattened envelope}\label{Disk Mass}

We derive the total envelope mass from the continuum flux density with the formula $M_{\rm tot}=[{F_\nu\,D^2}/B_\nu (T_{\rm dust})]{\it C_\nu}$ \citep{Hildebrand1983}, where $D=\,$11.11~kpc  is the distance to W49N \citep{Zhang2013} and $C_\nu$ is the emissivity of dust particles.
The value of $C_\nu$ at 226~GHz is extrapolated from its value at 400$\,\mu$m, $C_{400\,\mu\rm{m}}$, with the frequency dependence of $\nu^{-\beta}$.
We use $C_{400\,\mu\rm{m}}$\,=\,27~g\,cm$^{-3}$, which was obtained by \citet{Keene1982} under the assumption of gas to dust ratio of 100.

For $\nu$=226~GHz ($\lambda$=1,330 $\micron$) and $T_{\rm dust}\gg10$\,K, the above formula becomes 
\begin{equation}
M_{\rm tot}=10,170\,\times3.325^{\,\beta}\,\frac{F_\nu\,[{\rm Jy}]}{{T_{\rm dust}\,[{\rm K}]}}~{\rm M}_\odot.
\end{equation}
We will use $\beta=2$ for the spectral index of dust emissivity.
This is because the rotating structure is much larger ($\sim10,000$~au) and younger (\la\,10$^5$~yr) than the protoplanetary disks around low-mass stars, toward which smaller values of (0\,\la\,$\beta$\,\la\,1.5) are reported \citep[e.g.,][]{Beckwith1991}, and we may naturally assume that the optical properties of dust particles in the MCN-a envelope are similar to those of the interstellar dust particles.
Because of the uncertainty in $\beta$ and then in $C_{226~\rm GHz}$, which also bears the uncertainty in the gas to dust ratio, we estimate the overall uncertainty in the mass derived below to be a factor of three.

The dust temperature is empirically assumed to be 
\begin{equation}\label{T-R}
T_{\rm dust}=300\,(R~{\rm [au]}/1,700)^{-2/(4+\beta)}~{\rm K}
\end{equation}
so that it matches the rotation temperature $T_{\rm dust}$(1,700~au) = 300~K derived from CH$_{3}$CN and HNCO for $R\,\la\,$ 1,700~au with the approximate radial dependence of $R^{-2/(4+\beta)}$ suggested by numerical calculations \citep{Churchwell2002, Indebetouw2006}.
We then obtain $T_{\rm dust}=180$~K at the radius of $R=7,800$~au (0\,\farcs70).

The flux density 1.46~Jy integrated over the dust emission gives the total mass of 910~M$_\odot$ within the radius of 7,800~au (0\,\farcs70).
The flux density in the beam solid angle toward the emission peak is 84.3\,mJy (see Table~\ref{Table01}), which gives the mass of 32~M$_\odot$ contained within the radius of 1,700~au.
Note that these masses do not include the stellar mass and have an uncertainty of a factor of three.

\subsection{Mass distribution}\label{Mass Distribution}

\begin{figure}[htbp]
\includegraphics[bb= 50 20 600 500, scale=0.5]{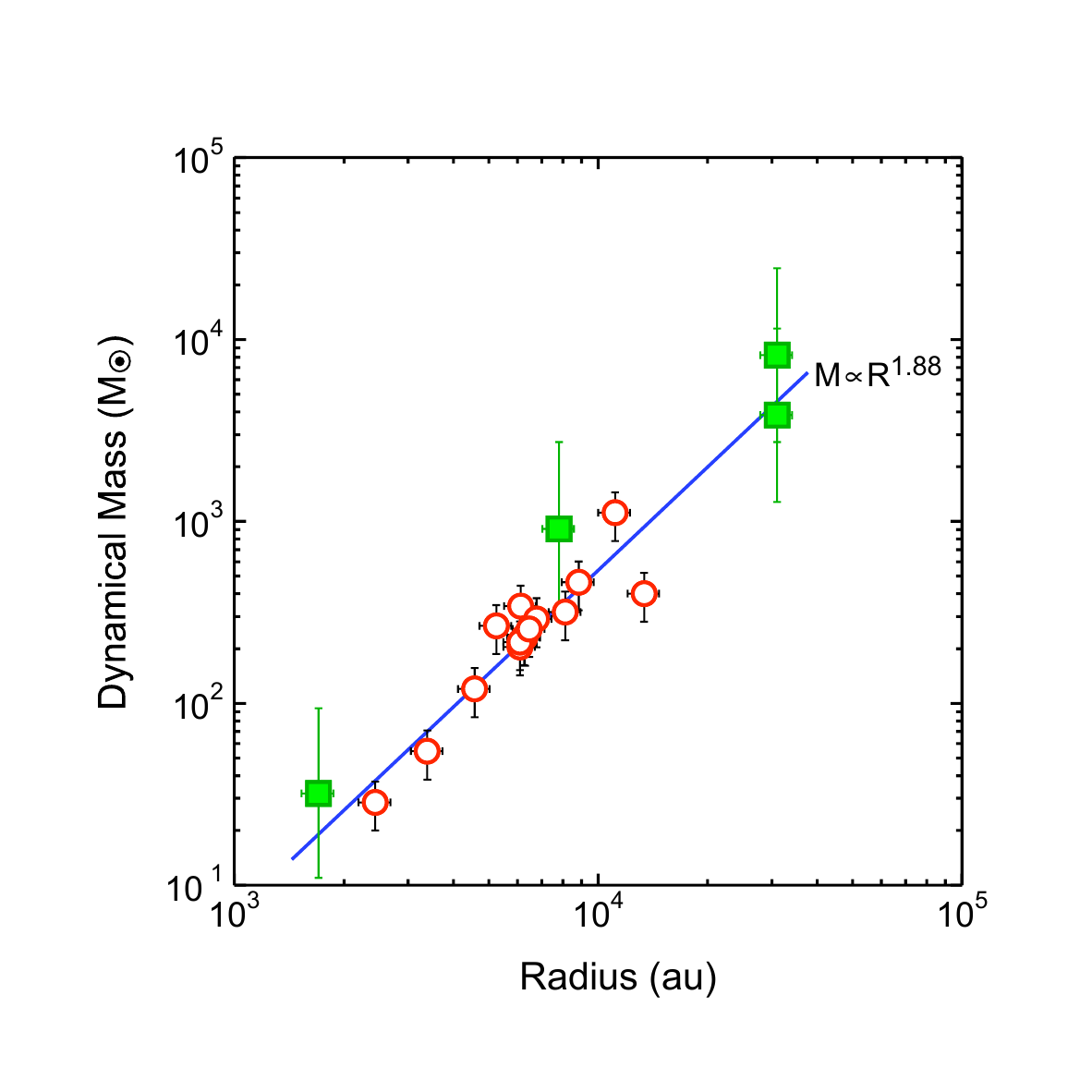}
\caption{Dynamical mass contained in the radius $R$ is shown in red open circles.
The blue line shows a power law fit.
The green squares at $R$\,= 1,700~au, 7,800~au, and 31,000~au indicate the mass within the respective radii estimated from the dust and C$^{18}$O emissions.
Note that the masses in green squares do not include the stellar mass, whose contribution should be significant for the data at $R$\,= 1,700~au.
}
\label{Fig18}
\end{figure}

In Figure~\ref{Fig18} we show a plot of the dynamical mass calculated by $M_{\rm dyn}=R\,V_{\rm rot}^2/G$ from each effective radius $R$ and inclination corrected rotation velocity $V_{\rm rot}$ listed in Table~\ref{Table02}.
The data points are fitted by a power law as $M_{\rm dyn}\propto R^{1.88\pm0.23}$.
If we give the coefficient for the specific power law index of 1.88, the relation is expressed as
\begin{equation}\label{Mdyn all}
M_{\rm dyn}=57.0^{+24.5}_{-17.1}\,(R\,[{\rm au}]/3,000)^{1.88}~{\rm M}_\odot.
\end{equation}
We did not include infall in calculating the dynamical mass, because it was not detected.
As we discuss in the next subsection, however, we expect that there should be infall with a velocity of up to $V_{\rm infall}\sim$2\,km\,s$^{-1}$.
If we take this effect into account, the additional contribution of $R\,V_{\rm infall}^2/G$ should be added to the dynamical mass.
This would increase it by 30\% at most.

The masses obtained from the dust continuum emission, 910~M$_\odot$ for $R\le7,800$~au and 32~M$_\odot$ for $R\le1,700$~au, are also plotted in green squares at the respective radii, with their uncertainties shown in error bars.
The two types of mass estimates, as well as their radial dependences, agree well with each other.
The good agreement suggests that the dynamical mass formula (\ref{Mdyn all}) well represents the mass distribution in MCN-a, and, in turn, supports the validity of the rotation curve that appears to have an opposite trend to the overall rotation velocity distribution in MCN-a.

The mass estimate 32~M$_\odot$ for $R\le1,700$~au does not include the stellar mass.
Assuming the central stellar mass to be 14--15\,M$_\odot$, corresponding to a field B0 star \citep[e.g.,][]{Hohle2010} and adding it to the mass derived from the dust emission, we obtain the total mass within 1,700~au to be 46--47~M$_\odot$.
The dynamical mass for $R\le1,700$~au calculated from the formula~(\ref{Mdyn all}) is $19.6^{+8.4}_{-5.9}$\,M$_\odot$, which is smaller than 46~M$_\odot$, but is within its uncertainty.

Our method of determining the effective radii and velocities has an error of $\sim$10\%, resulting in the error of $\sim$30\% in the dynamical mass for each plotted point.
The dynamical mass is directly derived from observed parameters and has a much smaller error than the mass derived from the dust and line fluxes based on various assumptions, although there is a systematic uncertainty that may lower all the data points of dynamical mass by $\sim$30\% caused by the uncertainty in the inclination angle of the envelope.
There is another uncertainty to raise the data points up to 30\% caused by our ignoring possible infall in calculating the dynamical mass. 

If we extrapolate the formula (\ref{Mdyn all}) to the radius of 1,000~au, we obtain $7.2^{+3.1}_{-2.2}$\,M$_\odot$, which now becomes a little smaller than the stellar mass, implying that the formula (\ref{Mdyn all}) and the velocity law (\ref{eq:rotation curve}) no longer hold at this small radius: the rotation velocity should be higher there in order to match the stellar mass.
This also means that the envelope mass has only minor contribution to the total mass at $R\,\la\,1,000$~au.
Although the envelope is massive and self-gravitating outside this radius, the stellar mass becomes dominant inside it and the envelope approaches a non self-gravitating Keplerian-like disk.
The decreasing trend of rotation velocity with decreasing radius should turn into increase, as is also supported by the compact high velocity molecular emission toward the center.
This is also consistent with the observed sizes of $\sim$1,000~au for Keplerian-like disks around O-type stars \citep{Johnston2015, Cesaroni2017, Maud2019, Zhang2019}. 

Let us examine the mass distribution at larger radii of the envelope.
The protostar, disk, and flattened envelope system of MCN-a is embedded in the HMC SiO-NE with a mass of $(1-2)\times10^4$~M$_\odot$ \citep{Miyawaki2022}.
We saw in \S\ref{C18O} that SiO-NE is well represented by the integrated emission of C$^{18}$O.
From the integrated intensity of the C$^{18}$O emission toward the center ($\int T_{\rm b}\,dV=531$~K\,km\,s$^{-1}$), its peak column density is calculated by 
\begin{equation}
N({\rm C}^{18}{\rm O})=1.06\times10^{14}\frac{\exp[-5.27/T_{\rm ex}]}{1-\exp[-10.56/T_{\rm ex}]}\int T_{\rm b}\,dV~~{\rm cm^{-2}},
\end{equation}
where $T_{\rm ex}$ is the excitation temperature and is assumed equal to the gas and dust temperature under the assumption of LTE.
This gives $N$(C$^{18}$O)=6.83$\times10^{17}$~cm$^{-2}$ with $T_{\rm ex}$=114~K given by the temperature distribution (\ref{T-R}).
We then obtain the total gas mass to be $(3.8-8.2)\times10^3$~M$_\odot$ by multiplying the effective  emission area ($0.19\times0.12\,(\pi/\ln\,2)~{\rm pc}^{2}$)  for an elliptical gaussian distribution of C$^{18}$O under the assumed abundance ratio of $X$(C$^{18}$O)= $(1.70-3.64)\,\times 10^{-7}$ with respect to H$_2$ \citep{Dickman1978, Frerking1982}.
Note that the total mass thus derived has an uncertainty of a factor of three mainly reflecting the uncertainty in the abundance ratio of C$^{18}$O.
The mass derived from the C$^{18}$O emission agrees with that of (1--2)$\times10^{4}$~M$_\odot$ derived from the SiO emission within the uncertainty.

If we extrapolate the dynamical mass formula (\ref{Mdyn all}) to the radius of 0.15~pc (31,000~au), we obtain $4,600^{+2,000}_{-1,400}$~M$_\odot$, which is well consistent with the above derived total mass of the entire HMC.
The mass dependence of $M(R)\propto R^{1.88\pm0.23}$ may thus be valid for larger radii up to the size of the entire HMC. 
Although the rotation law $V_{\rm rot}\propto R^{0.44\pm0.11}$ was derived for 2,400~au \la\,$R$\,\la\ 14,000~au and may be valid only for 2,400~au \la\,$R$\,\la\ 11,000~au, the dynamical mass derived from this law seems to be valid for 1,700~au \la\,$R$\,\la\ 31,000~au.
The mass-radius relation indicates that the density $\rho(R)$ depends on radius as $R^{-1.12\pm0.23}$, flatter than the singular isothermal case ($\rho\propto R^{-2}$) and observationally determined density laws with the power law indices of $-2.6$ to $-$1.5 \citep[e.g.,][]{Palau2014}.

\subsection{Accretion in the rotating envelope}\label{Disk accretion}

The derived dependence of mass on radius implies that the surface density of the flattened envelope varies as $\Sigma\propto R^{-0.12\pm0.23}$.
If we give the coefficient for the specific power law index of $-0.12$, the relation becomes 
\begin{equation}\label{eq:surface density}
\Sigma= 16.9^{+7.2}_{-5.1} \,(R\,[{\rm au}]/3,000)^{-0.12}$~g\,cm$^{-2}.
\end{equation}

Such a massive rotating structure should inevitably be unstable and, in fact, the average Toomre $Q$-value is calculated to be 
$Q=0.373\,(R~{\rm [au]}/3,000)^{-0.44}$ with the above surface density law and the assumed temperature of 200~K.
The $Q$-value is less than one for $R\,\ga\,300~{\rm au}$.
We infer that the rotating envelope at radii larger than 300--1,000~au is gravitationally unstable and form spiral arms and fragments, eventually accreting to the inner radii at a rate comparable to the free fall time scale \citep[e.g.,][]{AndreOliva2020}.

The free fall time scale of an object at $R=3,000$~au around a system of 57\,M$_\odot$ (see formula~(\ref{Mdyn all})) is 7,100~yr, which would give a mass accretion rate of $\sim8.0\times10^{-3}$\,M$_{\odot}\,yr^{-1}$ with an average inward velocity of 2.0\,km\,s$^{-1}$.
This magnitude of inward velocity, equal to the velocity resolution by chance, possibly justifies the negative detection of the inward motion.
The accretion rate could then be up to of order 10$^{-2}$\,M$_{\odot}\,yr^{-1}$ at $R=3,000$~ au in the envelope, consistent with that for massive star formation.
Higher velocity resolution molecular line observations would successfully detect the inward motion and determine the accretion rate in the envelope.

The high accretion rate is consistent with the absence of the broad-line H30$\alpha$ emission presented in Appendix~\ref{U-line}.
It implies that the disk in the vicinity (\la10~au) of the central star has not yet been sufficiently ionized.
One puzzle is that MCN-a has already formed a UCHII region (J$_{1}$) before developing an HCHII region, apparently not following the general understanding that UCHII regions are a later stage than HCHII regions.
One possibility may be that the 8.3 GHz flux is not entirely attributed to the thermal free-free emission because the UCHII region apparently does not have a rising, or at least flat, spectrum at centimeter wavelengths (see Figure~\ref{Fig02}).
In this case, the UCHII region may not be ionized chiefly by the stellar Lyman continuum photons and the photo-ionized UCHII region may have not yet developed.

There are pieces of evidence that the central star is more massive than B0 as we saw in \S\ref{Trot}.
Although the central (proto)star of MCN-a has a Lyman continuum photon luminosity comparable to a B0 star \citep{DePree1997}, the large size of the high temperature region (300~K at $R$~=~1,700~au), as well as the large visual extinction \citep{DeBuizer2021}, suggests that MCN-a intrinsically has a higher luminosity of order 10$^6$~L$_\odot$.
It would be difficult for a protostar to produce such a high luminosity only from accretion \citep[e.g.,][]{Hosokawa2009}, so most of the luminosity should originate from hydrogen burning.
If this is the true luminosity of the central star, only a few percent of its total luminosity is utilized for ionizing the ambient envelope.
It might be the case that only a tiny portion near the central star, such as the less dense polar regions evacuated by the outflow, is ionized in MCN-a, while the denser accretion disk absorbs most of the ionizing photons yet remaining still neutral because of the high accretion rate.
The true nature of the MCN-a UCHII region, utterly inconspicuous with negative spectral index at centimeter wavelengths, is yet to be known.
In any case, MCN-a is considered to be before the formation of an HCHII region, suggesting that it is in the earliest phase of massive star formation.
Absence of directly associated maser emission (CH$_{3}$OH class I and II, H$_{2}$O, SiO, and OH) often observed toward UCHII regions also supports its youth.

\section{Summary}\label{Summary}

\begin{figure}[htbp]
\includegraphics[bb= 0 120 600 680, scale=0.4]{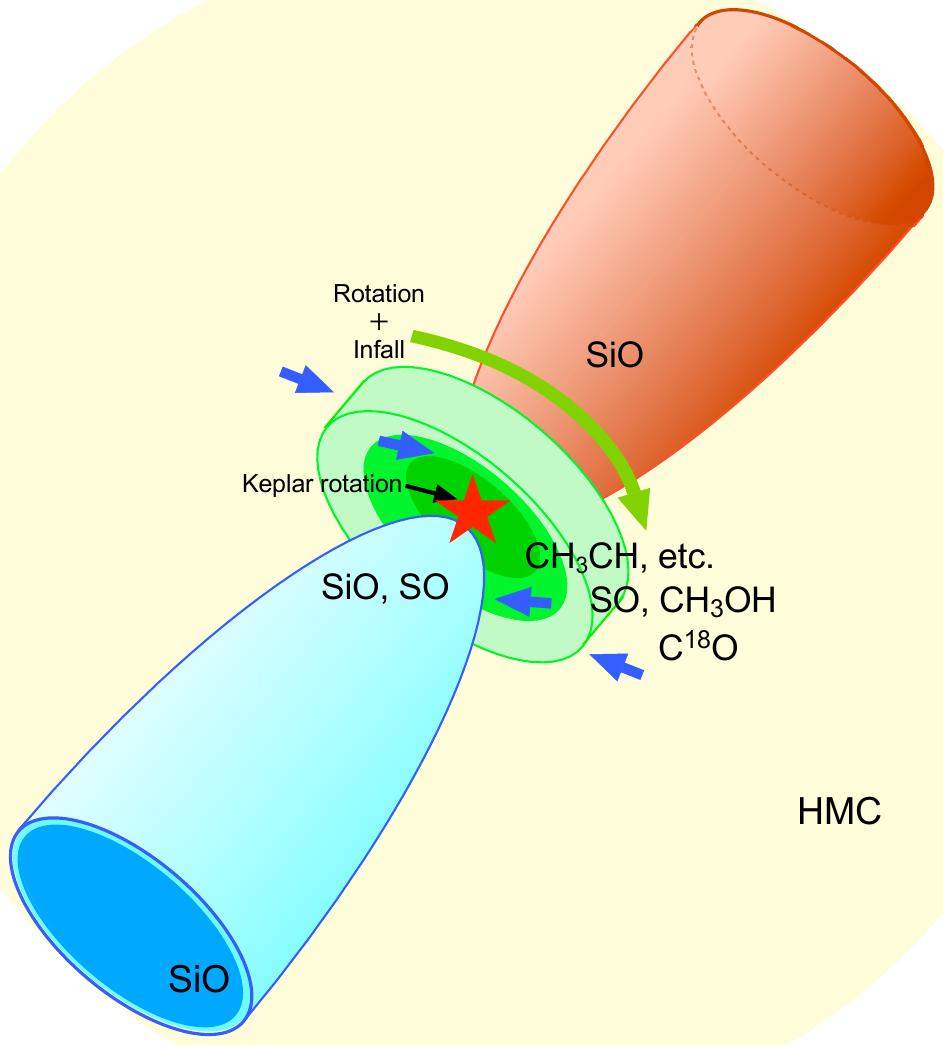}
\caption{A schematic diagram of the HMC W49N MCN-a and its rotating, self-gravitating, flattened envelope with a molecular outflow.
}
\label{Fig19}
\end{figure}

We have presented ALMA archival data for 219--235~GHz continuum and line observations toward the HMC W49N MCN-a carried out at an angular resolution of $\sim$0\farcs3 (3,300~au).
Our main results and conclusions are summarized as follows and illustrated in the schematic diagram in Figure~\ref{Fig19}.

\begin{enumerate}

\item 
The 226~GHz dust continuum emission shows an elongated structure of 1\farcs40~$\times$~0\farcs95 (deconvolved FWHM size) with a PA=43.5\degree\ perpendicular to the molecular outflow seen in the SiO and SO emissions. 
This structure is interpreted as a rotating flattened envelope (or torus) with a radius of 7,800~au and an inclination angle of 47.5\degree\ or larger.

\item
The velocity channel maps of CH$_3$CN ($J_K=12_3-11_3$), $^{13}$CS ($J=5-4$), HNCO ($J_{K_a,\,K_c}=10_{\,0,\,10}-9_{\,0,\,9}$), HC$_3$N ($J=24-23$), SO$_2$ ($J_{K_a,K_c}=28_{\,3,25}-28_{\,2,26}$), DCN ($J=5-4$), H$_2$CO ($J_K=3_{22}-2_{21}$), OCS ($J=19-18$), and CH$_3$OH ($J_K=4_{3}-3_{2}$) emissions show a consistent velocity gradient along the major axis of the envelope, manifesting its rotation.

\item
The SiO and SO emissions show that the outflow is blueshifted on the SE side and redshifted on the NW side, suggesting that the NW side of the flattened envelope is the near side.

\item
The C$^{18}$O emission is extended more than the dust envelope, still showing rotation around the outflow axis.
Its integrated intensity image has a size of 7\farcs00\,$\times$\,4\farcs44 (PA=173\degree), or 0.19~pc and 0.12~pc, giving the total mass of $(3.8-8.2)\times10^3$~M$_\odot$.
The mass and size agree well with those of SiO-NE, an HMC identified in a previous paper.

\item
The PV diagrams of various emission lines show velocity gradients along the major axis of the envelope as a result of rotation.
The magnitude of each velocity gradient is different, reflecting that each molecular line samples a specific radial region of the envelope rotating at a different velocity.

\item
For each PV emission, we derived the effective radius and rotation velocity by fitting a two dimensional gaussian.
Fitting the inclination corrected rotation velocity with a power law relation, we obtained a rotation curve as $V_{\rm rot}\propto R^{0.44\pm0.11}$ for 2,400~au $\le R \le$ 14,000~au.

\item
This rotation velocity law, increasing outward, is opposite to the overall trend of decreasing velocity from the stellar vicinity to the outer region of the hot core.
The opposite trend could be due to the pile-up of mass in the intermediate range of radius.

\item
Other than SiO and SO, which trace the outflow, the PV diagrams along the minor axis of the envelope do not show a clear velocity gradient indicative of infall.
Rotational motion is dominant in the envelope.

\item
Using multiple transitions of CH$_3$CN, HCNO and CH$_3$OH, we derived their rotation temperatures as $T_{\rm rot}$(CH$_3$CN) = 278\,$^{+39}_{-30}$~K, $T_{\rm rot}$(HNCO) = 297\,$^{+52}_{-39}$~K, and $T_{\rm rot}$(CH$_3$OH) = 154\,$^{+73}_{-37}$~K for $R\,\la$\,1,700~au.
The measured high temperatures at such a large radius suggests that the central source of MCN-a has an intrinsic bolometric luminosity of $\sim10^6$~L$_\odot$.

\item
The flux density of 1.46~Jy integrated over the entire dust continuum emission gives the total mass of 910\,M$_\odot$ (for $T_{\rm dust}=180$~K) for $R\le7,800$~au (0\,\farcs70) with an overall uncertainty of a factor of three.

\item
The flux density of 84.3\,mJy in the beam solid angle toward the dust emission peak gives the mass (excluding the central stellar mass) of 32\,M$_\odot$ (for $T_{\rm dust}=300$~K) within the radius of $R\le1,700$~au  (0\,\farcs15) with an overall uncertainty of a factor of three.

\item
The rotation curve derived from the velocity gradients of molecular emissions gives the dynamical mass formula as 
$M_{\rm dyn} = 57.0^{+24.5}_{-17.1}\,(R\,[{\rm au}]/3,000)^{1.88}$\,M$_\odot$ for 2,400~au $\le R \le$ 14,000~au, which is well consistent with the masses derived from the dust emission for $R\le7,800$~au and $R\le1,700$~au.

\item
We would obtain the dynamical mass within $R=1,000$~au to be $7.2^{+3.1}_{-2.2}$\,M$_\odot$, if we extrapolate the mass formula to the inner region of the envelope.
Because the dynamical mass is apparently smaller than the stellar mass, this means that the rotation law and mass formula no longer hold at this radius. 
This also impplies that the envelope may not be self-gravitating at $R\la1,000$~au and should become a Keplerian-like rotating disk there.

\item
By extrapolating the mass formula to the outer region at $R=0.15$~pc (31,000~au), we obtain the entire mass of the HMC to be $4,600^{+2,000}_{-1,400}$\,M$_\odot$, which agrees well with the mass of 3,800--8,200~M$_\odot$ estimated from the C$^{18}$O.
The mass formula may thus be valid for up to $R\sim$\,0.15~pc.

\item
The mass formula gives the density distribution as $\rho(R)\propto R^{-1.12\pm0.23}$, shallower than those previously reported.

\item
The differential rotation of the flattened envelope and its surface density give the Toomre $Q$-value less than one for $R\,\ga\,300$~au.
This implies that the rotating envelope at $R\,\ga\,(300-1,000)$~au is gravitationally unstable and form spiral arms and fragments, allowing the gas and dust to accrete at a rate comparable to the free-fall timescale of 7,100~yr.
The mass accretion rate would then be of order 10$^{-2}$\,M$_\odot$\,yr$^{-1}$ at $R = 3,000$~au.

\item
These results, together with the negative detection of the broad-line H30$\alpha$ emission, suggest that W49N MCN-a is in a very early phase of massive star formation, when a HCHII region has not yet fully developed around an accreting protostar with a mass of 14--15\,M$_\odot$.

\item
The current data has revealed the structure and kinematics of an HMC at its intermediate radii between the Keplerian-like disk and entire gas clump, providing an example of how gas accretes inside the HMC in the earliest phase of massive star formation.

\end{enumerate}

\begin{ack} 
We are grateful to the anonymous referee and Dr. K. E. I. Tanaka for useful comments, which have greatly improved this paper.
ALMA is a partnership of ESO (representing its member states), NSF (USA) and NINS (Japan), together with NRC (Canada), NSC and ASIAA (Taiwan), and KASI (Republic of Korea), in cooperation with the Republic of Chile. 
The Joint ALMA Observatory is operated by ESO, AUI/NRAO and NAOJ.
We used the ALMA archival data \#2016.1.00620.S: PI A. Ginsburg. 
\end{ack}

\begin{appendix}\label{APPENDIX}

\section{Channel maps of other lines}\label{Channel-maps}

\subsection{$^{13}$CS emission}\label{13CS}
\begin{figure*}[htbp]
\includegraphics*[bb= 0 210 700 620, scale=0.75]{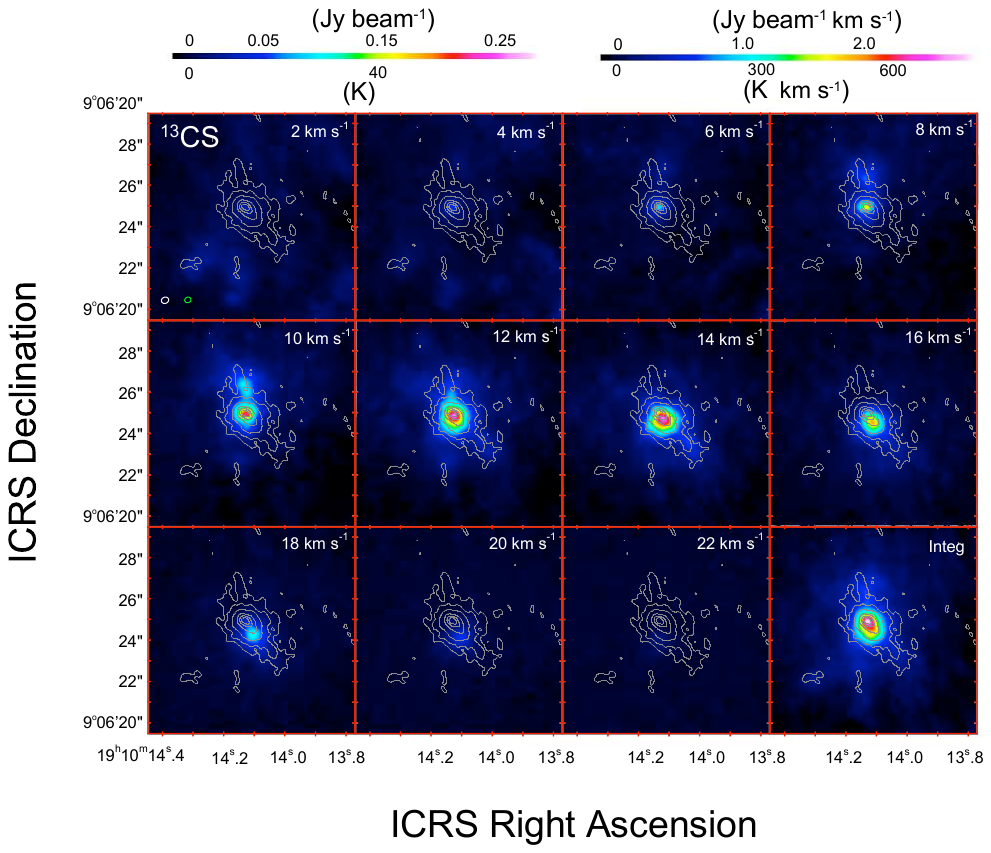}
\caption{Same as Fig.~\ref{Fig06}, but for the $^{13}$CS ($J=5-4$) line.
\label{Fig20}}
\end{figure*}

Figure~\ref{Fig20} shows the velocity channel maps of the $^{13}$CS ($J=5-4$) line.
The emission is detected at velocities from $V_{\rm LSR}$ =\,4\, km\,s$^{-1}$ to 22\, km\,s$^{-1}$.
Most of the emission arises within the 20\% level contour of the continuum emission.
The emission peak moves from the NE to SW with respect to the continuum peak as the radial velocity increases from $V_{\rm LSR}$ =\,6\, km\,s$^{-1}$ to 14\, km\,s$^{-1}$.
The line and continuum peaks coincide with each other at 10\, km\,s$^{-1}$.
Weaker emission features are seen at 1$''$--1\farcs5 N of the peak at 10\,km\,s$^{-1}$.

\subsection{HNCO emission}
\begin{figure*}[htbp]
\includegraphics*[bb= 0 210 700 640, scale=0.7]{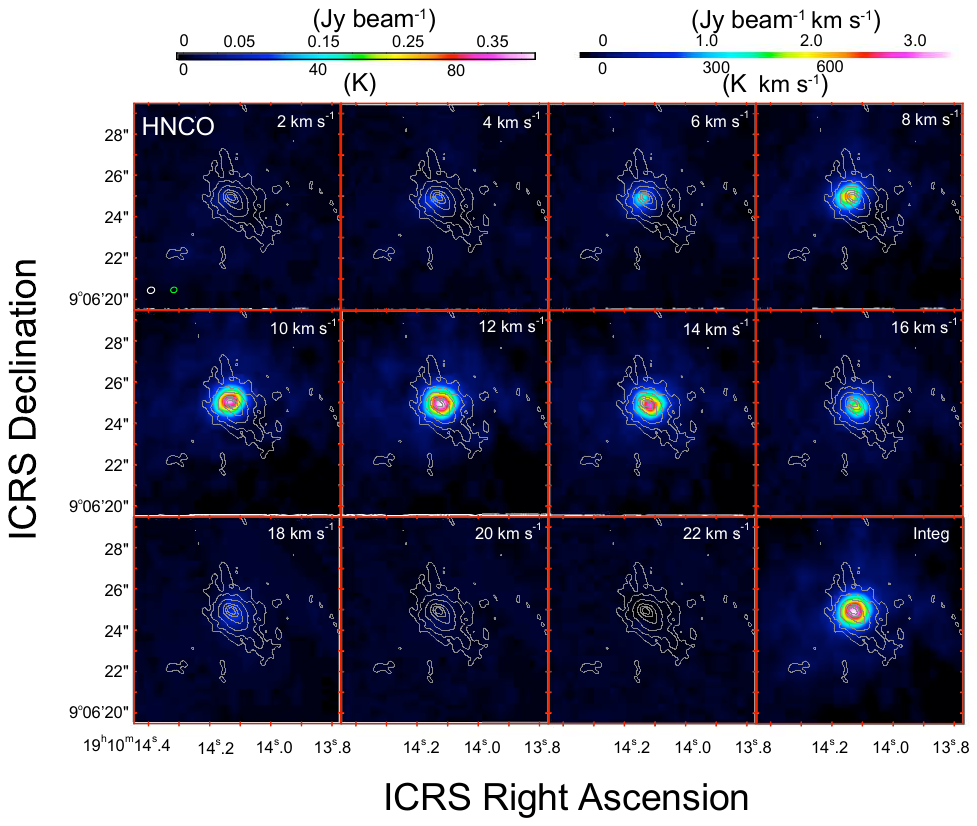}
\caption{Same as Fig.~\ref{Fig06}, but for the HNCO ($J_{K_a,\,K_c}=10_{\,0,\,10}-9_{\,0,\,9}$) line.
\label{Fig21}}
\end{figure*}

Figure~\ref{Fig21} shows the velocity channel maps of the HNCO ($J_{K_a,\,K_c}=10_{\,0,\,10}-9_{\,0,\,9}$) line.
The emission is detected at velocities from $V_{\rm LSR}$ =\,2\, km\,s$^{-1}$ to 18\, km\,s$^{-1}$.
The dominant part of the emission arises within the 20\% level contour of the continuum emission.
Similar to the case of CH$_3$CN, the emission peak moves from the NE to SW of the continuum peak as the radial velocity increases from $V_{\rm LSR}$ =\,8\, km\,s$^{-1}$ to 16\, km\,s$^{-1}$.
The line and continuum peaks coincide with each other at 12\, km\,s$^{-1}$.

\subsection{HC$_3$N emission}
\begin{figure*}[htbp]
\includegraphics*[bb= 0 210 700 640, scale=0.7]{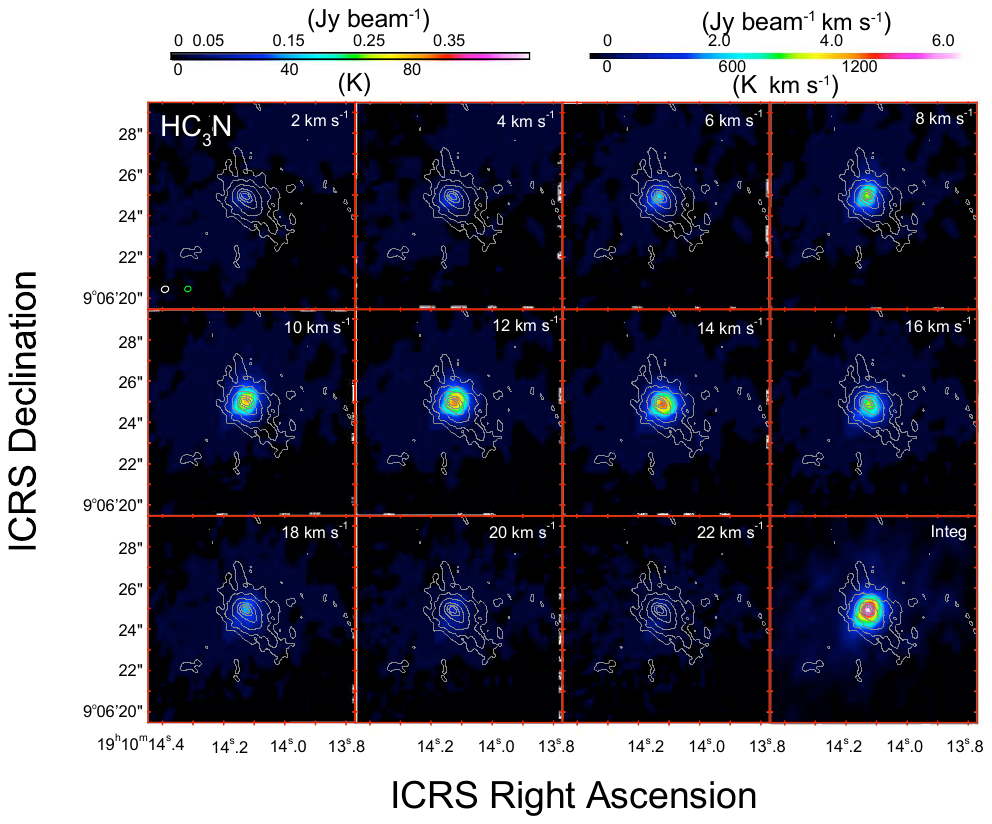}
\caption{Same as Fig.~\ref{Fig06}, but for the HC$_3$N ($J=24-23$) line.
\label{Fig22}}
\end{figure*}

Figure~\ref{Fig22} shows the velocity channel maps of the HC$_3$N ($J=24-23$) line.
The emission is detected at velocities from $V_{\rm LSR}$ =\,2\, km\,s$^{-1}$ to 22\, km\,s$^{-1}$.
The dominant part of the emission arises within the 20\% level contour of the continuum emission.
Similar to the case of CH$_3$CN, the emission peak moves from the NE to SW of the continuum peak as the radial velocity increases from $V_{\rm LSR}$ =\,8\, km\,s$^{-1}$ to 16\, km\,s$^{-1}$.
The line emission peak is located a little ($\sim$0\farcs3) NW of the continuum peak at 10 and 12\, km\,s$^{-1}$.

\subsection{SO$_2$ emission}
\begin{figure*}[htbp]
\includegraphics*[bb= 0 210 700 640, scale=0.7]{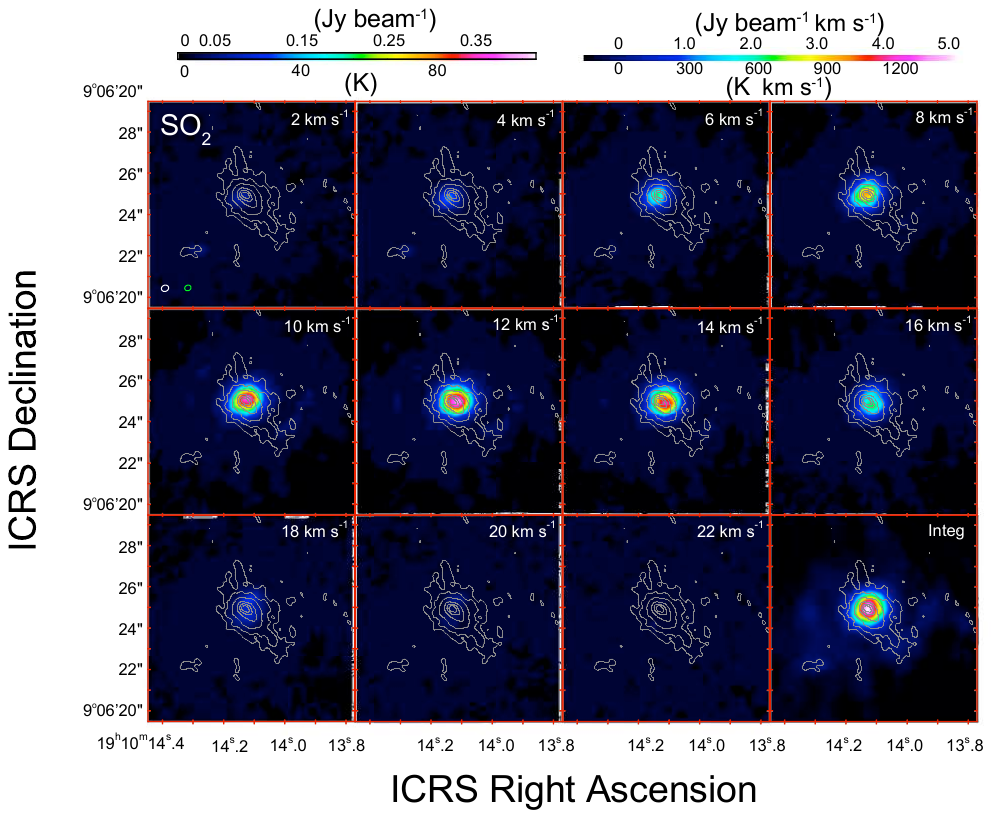}
\caption{Same as Fig.~\ref{Fig06}, but for the SO$_2$ ($J_{K_a,K_c}=28_{3,25}-28_{2,26}$) line.
\label{Fig23}}
\end{figure*}

Figure~\ref{Fig23} shows the velocity channel maps of the SO$_2$ ($J_{K_a,K_c}=28_{3,25}-28_{2,26}$) line.
The emission is detected at velocities from $V_{\rm LSR}$ =\,2\, km\,s$^{-1}$ to 20\, km\,s$^{-1}$.
The dominant part of the emission arises within the 20\% level contour of the continuum emission.
Similar to the case of CH$_3$CN, the emission peak moves from the NE to SW of the continuum peak as the radial velocity increases from $V_{\rm LSR}$ =\,8\, km\,s$^{-1}$ to 16\, km\,s$^{-1}$.
The peak of the line emission is shifted a little ($\sim$0\farcs3) to the NE of the continuum peak at 10\, km\,s$^{-1}$, where the two peaks are closest to each other.

\subsection{DCN emission}
\begin{figure*}[htbp]
\includegraphics*[bb= 0 210 700 640, scale=0.7]{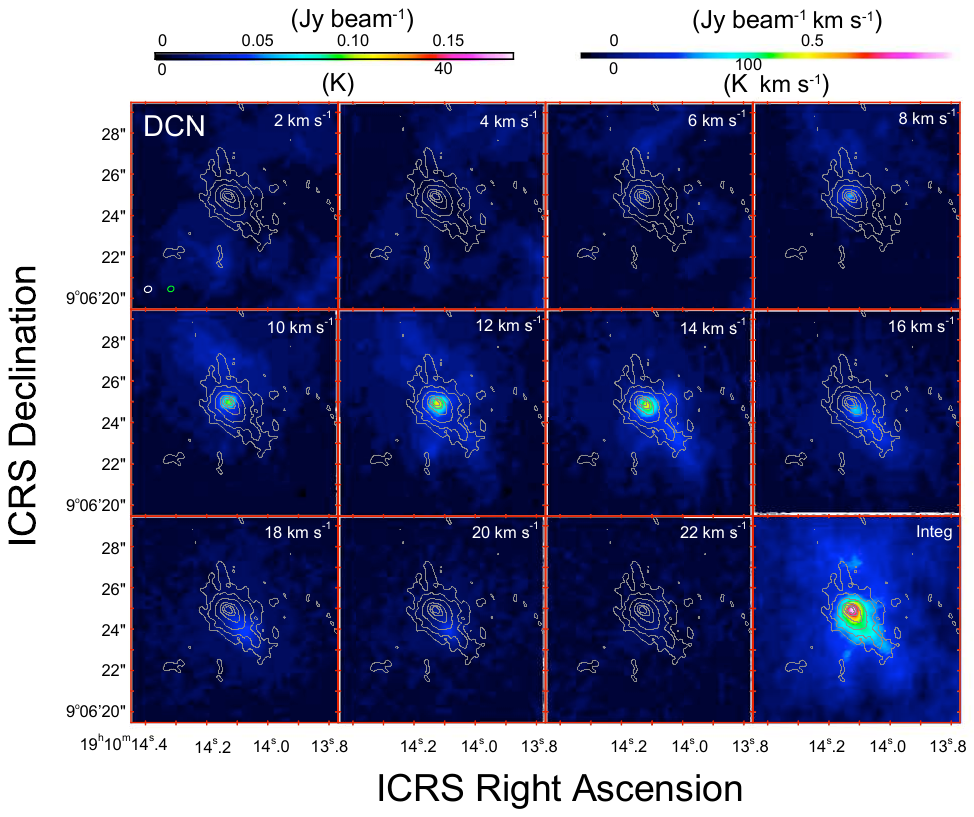}
\caption{Same as Fig.~\ref{Fig06}, but for the DCN ($J=5-4$) line.
\label{Fig24}}
\end{figure*}

Figure~\ref{Fig24} shows the velocity channel maps of the DCN ($J=5-4$) line.
Superposed on the weak extended emission, the feature associated with the dust emission is unambiguously detected at velocities from $V_{\rm LSR}$ =\,8\, km\,s$^{-1}$ to 20\, km\,s$^{-1}$.
The dominant part of the emission arises within the 40\% level contour of the continuum emission at 8\, km\,s$^{-1}\le\,V_{\rm LSR}\le$16\, km\,s$^{-1}$.
Similar to the case of CH$_3$CN, the emission peak moves from the NE to SW of the continuum peak as the radial velocity increases from $V_{\rm LSR}$ =\,8\, km\,s$^{-1}$ to 16\, km\,s$^{-1}$.
The line and continuum peaks coincide with each other within the beam size at 10--12\, km\,s$^{-1}$.

\subsection{H$_2$CO emission}
\begin{figure*}[htbp]
\includegraphics*[bb= 0 210 700 640, scale=0.7]{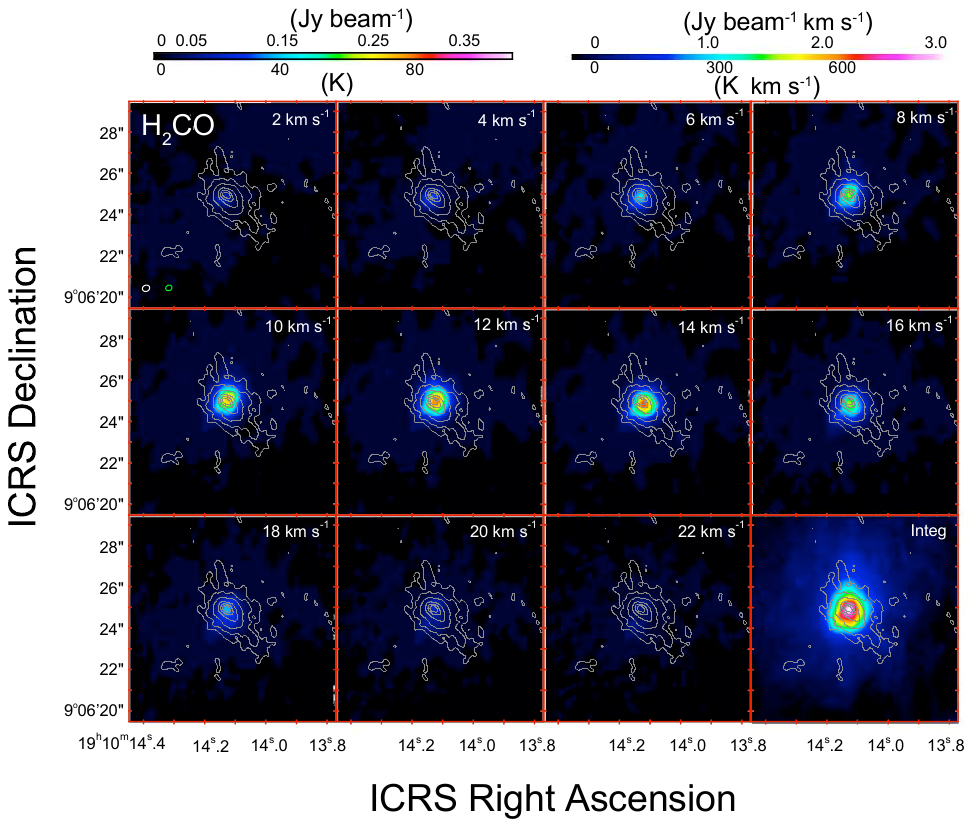}
\caption{Same as Fig.~\ref{Fig06}, but for the H$_2$CO  ($J_K=3_{22}-2_{21}$)  line.
\label{Fig25}}
\end{figure*}

Figure~\ref{Fig25} shows the velocity channel maps of the H$_2$CO  ($J_K=3_{22}-2_{21}$) line.
The emission is detected at velocities from $V_{\rm LSR}$ =\,2\, km\,s$^{-1}$ to 22\, km\,s$^{-1}$.
The dominant part of the emission arises within the 40\% level contour of the dust continuum emission.
Similar to the case of CH$_3$CN, the emission peak moves, although the positional shift is subtle, from the NE to SW across the continuum peak as the radial velocity increases from $V_{\rm LSR}$ =\,8\, km\,s$^{-1}$ to 16\, km\,s$^{-1}$.
The emission peak tends to ``return'' to the continuum peak at highly blueshifted (2\,km\,s$^{-1}$) and redshifted (18\,km\,s$^{-1}$) velocities, indicating that the inner part of the envelope rotates at a higher velocity.
This may be better seen in the position-velocity diagrams (Figures~\ref{Fig11} and \ref{Fig12}), although the emission features at $V_{\rm LSR}\,\ga\,25$\,km\,s$^{-1}$ there are not part of the H$_{2}$CO line.
The line and continuum peaks occur almost at the same position at 10\, km\,s$^{-1}$, but the line peak is shifted a little ($\sim$0\farcs3) to the NE of the continuum peak.

\subsection{OCS emission}
\begin{figure*}[htbp]
\includegraphics*[bb= 0 210 700 640, scale=0.7]{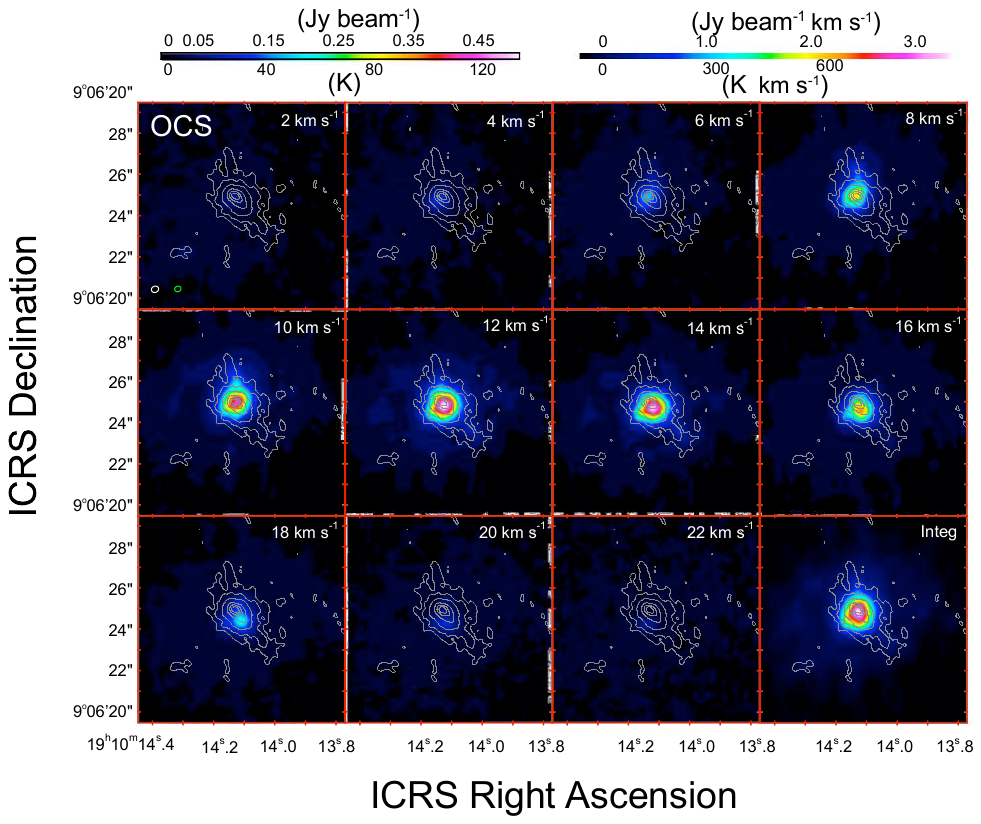}
\caption{Same as Fig.~\ref{Fig06}, but for the OCS ($J=19-18$) line.
\label{Fig26}}
\end{figure*}

Figure~\ref{Fig26} shows the velocity channel maps of the OCS ($J=19-18$) line.
The emission is detected at velocities from $V_{\rm LSR}$ =\,2\, km\,s$^{-1}$ to 22\, km\,s$^{-1}$.
The dominant part of the emission arises within the 20\% level contour of the continuum emission.
Similar to the case of CH$_3$CN, the emission peak moves from the NE to SW across the continuum peak as the radial velocity increases from $V_{\rm LSR}$ =\,8\, km\,s$^{-1}$ to 16\, km\,s$^{-1}$.
The line and continuum peaks coincide with each other at 10 and 12\, km\,s$^{-1}$.
A weaker emission feature is seen extended to the north from the peak at 8 and 10\, km\,s$^{-1}$.

\subsection{CH$_3$OH emission}
\begin{figure*}[htbp]
\includegraphics*[bb= 0 210 700 640, scale=0.7]{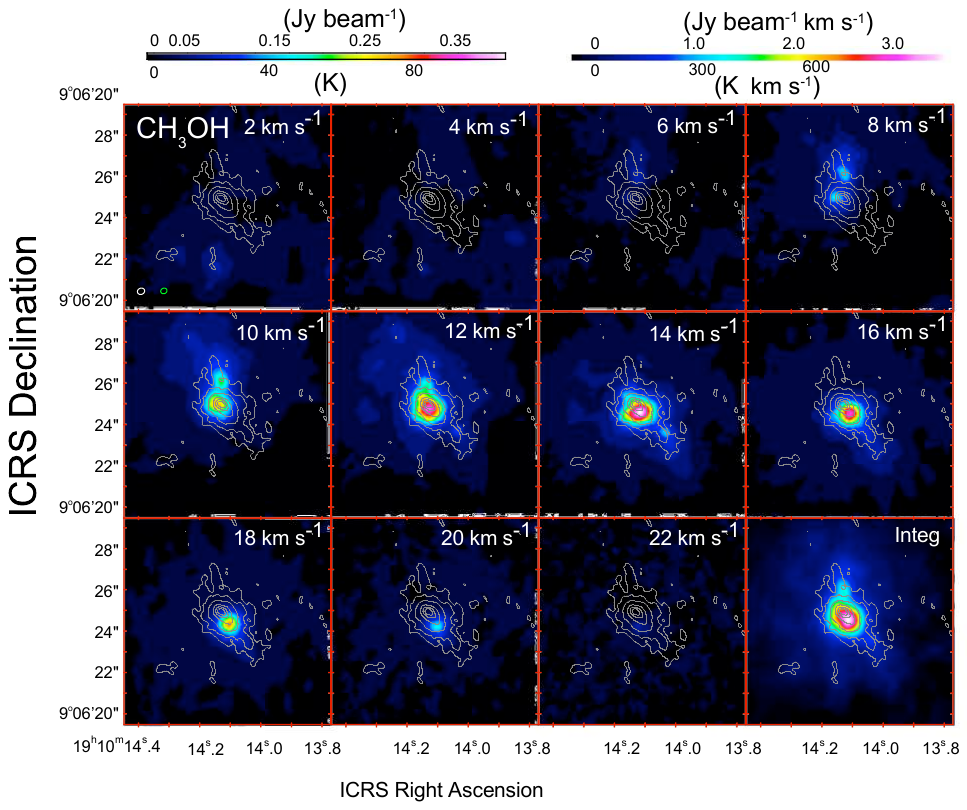}
\caption{Same as Fig.~\ref{Fig06}, but for the  CH$_3$OH ($J_K=4_{3}-3_{2}$) line.
\label{Fig27}}
\end{figure*}

Figure~\ref{Fig27} shows the velocity channel maps of the CH$_3$OH ($J_K=4_{3}-3_{2}$) line.
The emission is detected at velocities from $V_{\rm LSR}$ =\,8\, km\,s$^{-1}$ to 20\, km\,s$^{-1}$.
The main emission associated with the envelope exhibits a shift of peak positions from NE to SW as the radial velocity increases from $V_{\rm LSR}$ =\,8\, km\,s$^{-1}$ to 16\, km\,s$^{-1}$.
The line emission peak is slightly (\la\,0\farcs3) shifted to the south from the continuum peak at 12\, km\,s$^{-1}$, where the two peaks are closest from each other.
Other than the emission associated with the flattened envelope, a feature extended to the north from the continuum peak is seen at 8, 10 , and 12\, km\,s$^{-1}$, where a secondary peak is visible at $\sim$1\farcs2 north of the continuum peak.

\section{Unidentified line at 231.900~GHz}\label{U-line}

\begin{figure}[htbp]
\includegraphics[bb= 100 200 600 700, scale=0.55]{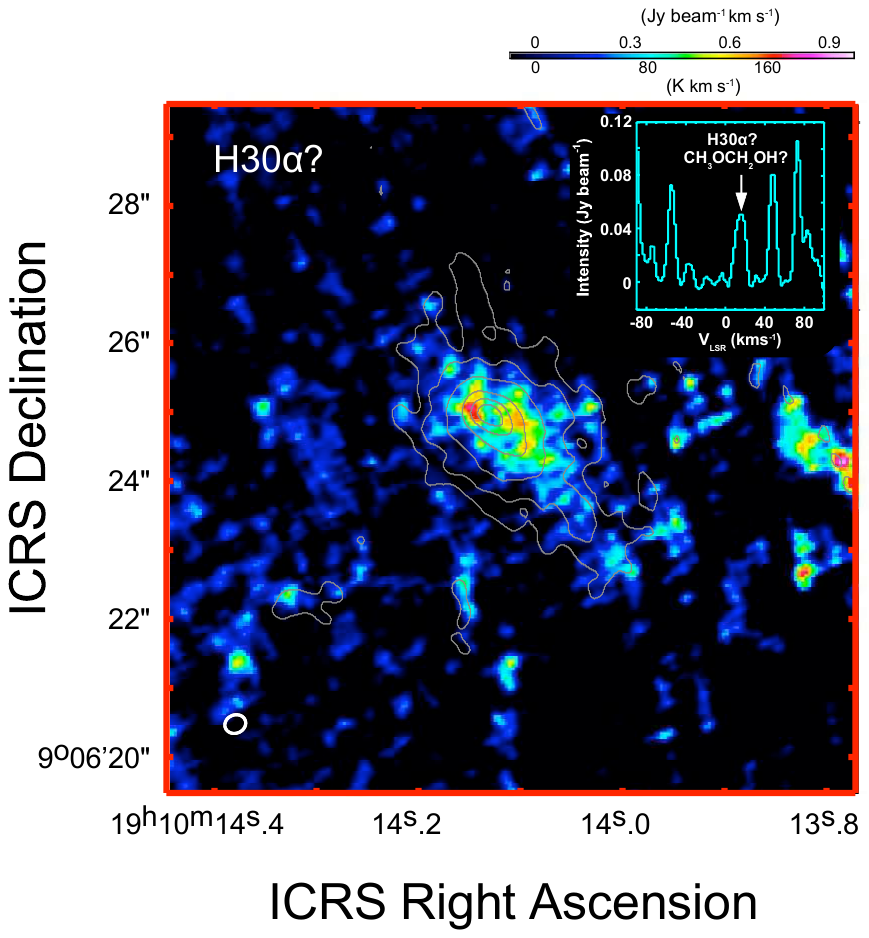}
\caption{Integrated intensity map of the emission feature at $\nu=\,$231.900~GHz, superimposed on the contours of the continuum emission at 5, 10, 20, 40, 60, and 80~\% levels of the peak brightness.
The inset shows the line profile toward the continuum peak with the horizontal axis being the LSR velocity with respect to the rest frequency of H30$\alpha$.
Integration was carried out from 0 to 30 km\,s$^{-1}$. 
}
\label{Fig28}
\end{figure}

There is an emission feature at the nominal frequency of the H30$\alpha$ recombination line ($\nu =$ 231.90092784~GHz).
Figure~\ref{Fig28} shows the integrated intensity map of the line.
The emission coincides with the flattened envelope, but is not centrally peaked.
We did not identify it as H30$\alpha$ because its line width (12~km\,s$^{-1}$) is similar to the other molecular lines and is significantly narrower than those of hydrogen recombination lines emitted by HCHII regions ($\Delta V$\,\ga\,40~km\,s$^{-1}$).
We suspect that this feature arises from CH$_3$OCH$_2$OH ($\nu =$ 231.90059920~GHz).

\section{Determination of the ridge line}\label{Ridge line}

\begin{figure}[htbp]
\includegraphics[bb= 150 400 600 700, scale=0.85]{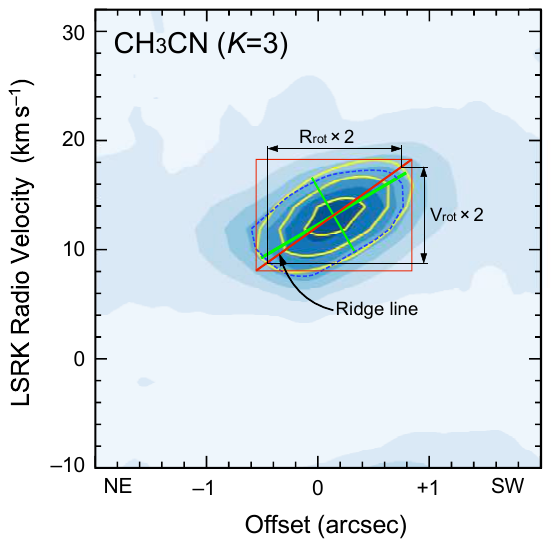}
\caption{Method of determining the ridge line is illustrated.
Bluish color map shows the PV diagram of CH$_{3}$CN ($J_K=12_3-11_3$) line.
The three yellow contours represent an elliptical gaussian fit to the data at the 50, 70, and 90\% levels of the peak brightness.
The ridge line is defined by the diagonal of the red rectangle that circumscribes the 50\% contour of the fitted ellipse.
The blue dotted contour indicates the 50\% level of the original PV data.
The two green lines orthogonal to each other are the major and minor axes of the fitted ellipse.
}
\label{Fig29}
\end{figure}

This section describes how we determine the ``ridge line" for a PV diagram.
Each PV data, consisting of 81 pixels $\times$ 22 pixels (4$''$ $  \times$ 42~km\,s$^{-1}$), is fitted by an elliptical gaussian using the pixels that have more than 50\% of the peak brightness.
The ridge line is defined as the diagonal of the rectangle circumscribing the 50\% contour of the fitted elliptical gaussian.
This is visually shown in Figure~\ref{Fig29} for the PV diagram of CH$_{3}$CN as an example.
The length of the ridge line defined on both sides by the 50\% contour of the fitted ellipse  provides (twice the value of) Rotation Radius $R$ when projected on to the positional axis.
Its projection on to the velocity axis gives (twice the value of) $V_{\rm rot}\,\sin i$.

Note that the ridge line is generally different from the major axis of the fitted ellipse, shown as a green line in Figure~\ref{Fig29}.
The inclination of the major axis of the ellipse measured in unit of velocity per arc second varies when positional and/or velocity axes are expanded or reduced independently, which makes the major axis not a good indicator of the velocity gradient.
In other words, the major axis of an ellipse is no longer the major axis of the same ellipse after its aspect ratio is varied.
On the other hand, the inclination (velocity per arc second) of the diagonal circumscribing the fitted ellipse does not change even when the positional and/or velocity axes are independently scaled.
Thus, whatever the aspect ratio of a PV diagram is, the diagonal always gives the same value of inclination in unit of velocity per arc second.

\end{appendix}


\end{document}